\documentclass[aps, prb, superscriptaddress,amsmath,amssymb,twocolumn, longbibliography]{revtex4-2}
\usepackage{graphicx}
\usepackage{subfigure}
\usepackage{adjustbox}
\usepackage{bm}
\usepackage{color}
\usepackage{braket}
\usepackage{standalone}
\usepackage{multirow}
\usepackage{tikz}
\usepackage{mathrsfs}
\usepackage{dsfont}
\usepackage{comment}
\usepackage{amsmath}
\usepackage{float}
\usepackage{enumitem}
\usepackage[colorlinks,bookmarks=true,citecolor=blue,linkcolor=red,urlcolor=blue]{hyperref}
\usepackage{cleveref}
\usepackage{orcidlink}
\usepackage{appendix}
\usepackage{graphicx}
\usepackage{caption}

\newcommand{\bK}{{\mathbf K}}
\newcommand{\bKp}{{\mathbf K^{\prime}}}

\newcommand{\bn}{{\mathbf n}}
\newcommand{\bs}{{\mathbf s}}

\newcommand{\bU}{{\mathbf U}}
\newcommand{\btau}{\bm{\tau}}
\newcommand{\vecr}{\vec{r}\,}
\newcommand{\vecq}{\vec{q}\,}

\newcommand{\vecqs}{\vecq_{\!*}}
\newcommand{\cdo}{c^{\dagger}}
\newcommand{\cao}{c^{ }}

\newcommand{\bea}{\begin{eqnarray*}}
\newcommand{\eea}{\end{eqnarray*}}
\newcommand{\bean}{\begin{eqnarray}}
\newcommand{\eean}{\end{eqnarray}}
\newcommand{\nn}{\nonumber\\}

\newcommand{\upa}{\uparrow}
\newcommand{\dwa}{\downarrow}

\begin{document}

\title{Fractional quantum Hall coexistence phases in  higher Landau levels of graphene}
\author{Jincheng An}
\email{Jincheng.An@uky.edu}
\affiliation{Department of Physics and Astronomy, University of Kentucky, Lexington, KY 40506, USA}
\author{Ajit C. Balram\orcidlink{0000-0002-8087-6015}}
\email{cb.ajit@gmail.com}
\affiliation{Institute of Mathematical Sciences, CIT Campus, Chennai, 600113, India}
\affiliation{Homi Bhabha National Institute, Training School Complex, Anushaktinagar, Mumbai 400094, India}
\author{Udit Khanna\orcidlink{0000-0002-3664-4305}}
\email{udit.khanna.10@gmail.com}
\affiliation{Department of Physics, Bar-Ilan University, Ramat Gan 52900, Israel}
\author{Ganpathy Murthy\orcidlink{0000-0001-8047-6241}}
\email{murthy@g.uky.edu}
\affiliation{Department of Physics and Astronomy, University of Kentucky, Lexington, KY 40506, USA}	
\

\begin{abstract}
Monolayer graphene under a strong magnetic field near charge neutrality manifests the integer and fractional quantum Hall effects. Since only some of the four spin/valley flavors available to the electrons in each Landau level manifold are filled, they also exhibit spontaneous symmetry breaking the in spin/valley sector, a phenomenon known as quantum Hall ferromagnetism. In this work, we study quantum Hall ferromagnets in the higher Landau level manifolds of monolayer graphene and show that there is an even richer set of symmetry-broken phases than in the lowest Landau level manifold. Specifically, both valley polarized and valley equatorial (where the occupied Landau levels are in an equal superposition of both valleys) ferromagnets, antiferromagnets, and canted antiferromagnets are found. Several types of spin valley entangled phases are found, all of which manifest the simultaneous spontaneous symmetry breaking of both magnetic and lattice symmetries.   
\end{abstract}
\maketitle

\section{Introduction}
\label{sec:Intro}

The integer quantum Hall effect (IQHE), discovered in two-dimensional electron gases (2DEGs) hosted in semiconductor heterostructures in 1980~\cite{IQHE_Discovery_1980}, is the simplest example of a topological insulator~\cite{TIs_RevModPhys.82.3045}. In IQH  states, electrons fill an integer number of Landau levels (LLs), leading to charge incompressibility in the bulk.   All electrical conduction occurs at the edges via gapless chiral boundary modes, which are stable against disorder. Soon after the discovery of the IQHE, the fractional quantum Hall effect (FQHE)~\cite{FQHE_Discovery_1982} was also found and understood to be stabilized by Coulomb interactions~\cite{Laughlin_1983}. These are also incompressible states that possess quasiparticle excitations with fractional charge and statistics~\cite{Wilczek82, Arovas84, Halperin84}. 

Interaction effects are important not only for the FQHE but also at integer fillings, where internal symmetries such as spin/valley rotations can be spontaneously broken. Such states are known as quantum Hall ferromagnets and can manifest topological excitations of the spin/valley order parameter known as skyrmions~\cite{Shivaji_Skyrmion}.

Two decades ago, an entirely new platform for the QHE was found in graphene~\cite{Novoselov_etal_2004, Novoselov_et_al_2005, Zhang_Kim_2005}, which is a two-dimensional honeycomb lattice of Carbon atoms with two atoms per unit cell on the sublattices $A$ and $B$. There are many variants of graphene with different numbers of layers and different stacking arrangements but in this work, we will focus on monolayer graphene. At zero magnetic field, graphene has two Dirac band touchings at the corners of the Brillouin Zone ($\bf K$ and ${\bf K}'$), near which the low-energy effective model is the two-dimensional relativistic Dirac equation. At charge neutrality, the chemical potential is at the Dirac points~\cite{neto2009:rmp}. Upon the application of a magnetic field ${\bf B_{\perp}}$ perpendicular to the graphene plane, the low-energy spectrum breaks up into a set of particle-hole symmetric LL manifolds indexed by an integer $n$ which can take values $n{=}0,{\pm}1,{\pm}2{\dots}$ with energy $E_n{=}{\rm sgn}(n)(\hbar v_F/\ell)\sqrt{|n|}$, where $\ell{=}\sqrt{\hbar/(eB_{\perp})}$ is the magnetic length at magnetic field strength $B_{\perp}{=}|{\bf B_{\perp}}|$ and $v_{F}$ is the Fermi speed. Each manifold has four (nearly) degenerate LLs after including the spin and valley degrees of freedom. The Zeeman coupling is typically the smallest scale in the problem because $v_F {\approx} 10^6$ m/s is quite large. The $n{=}0$ manifold of LLs, called the zero LLs (ZLLs), is special because it is an equal superposition of particle and hole states. Consequently, near a generic edge, the two valleys mix and produce one mode that has a particle-like dispersion, while the other has a hole-like dispersion~\cite{Graphene_Edge_Herb_2006}. The ZLL also has the property of valley-sublattice locking, where the wave functions in the ${\bf K}$ valley have support only on the $A$ sites, while the wave functions in the ${\bf K}'$ valley have support only on the $B$ sublattice. The wave functions in the $|n|{\geq}1$ higher LL manifolds are equal superpositions on the two sublattices. This fact is important when the sublattice symmetry is explicitly broken, for example, by the substrate potential of the encapsulating hexagonal Boron Nitride (hBN). 

Due to the 4-fold (near) degeneracy, the partially filled LL manifolds in graphene are prime candidates to display quantum Hall ferromagnetism (QHFM)~\cite{Shivaji_Skyrmion, QHFM_Moon_etal_1995}. The filling factor is normalized such that at charge neutrality i.e., half-filling, $\nu{=}0$ while the fully filled ZLL manifold has $\nu{=}2$. Many quantum Hall states, both integer and fractional, have been seen experimentally in monolayer graphene in the $n{=}0,1,2,3$ manifolds~\cite{Andrei_FQH2009, Bolotin_FQH2009, Hone_Multicomp_MGL2011, Hofstadter_Ashoori2013, GoldhaberGordon2015, Even_Den_Young2018, Kim19}. 

Theoretically, the nature of the ground state at $\nu{=}0$ was first addressed in Refs.~\cite{abanin2006:nu0, Brey_Fertig_2006}, where only the fully spin/valley [$SU(4)$] symmetric long-range Coulomb interactions were included. Two of the four LLs in the ZLL manifold must be filled, with the question being which two orthogonal linear combinations of the four available LLs are occupied. The prediction was that the ground state would be a spin ferromagnet~\cite{abanin2006:nu0, Brey_Fertig_2006}, with the valley being unpolarized. This state has a pair of counterpropagating edge modes with opposite spin which cannot backscatter due to spin conservation, which means this is a quantum spin Hall state~\cite{Kane_Mele_2DTI_PhysRevLett.95.146802}, with a two-terminal conductance of $2e^2/h$. However, experimentally, the state at purely perpendicular ${\bf B_{\perp}}$ was found to be a vanilla insulator with a vanishing two-terminal conductance~\cite{young2014:nu0}. When the Zeeman coupling is increased by adding a parallel field, the two-terminal conductance rises and eventually approaches $2e^2/h$~\cite{young2014:nu0}. Based on a large body of work~\cite{alicea2006:gqhe, KYang_SU4_Skyrmion_2006, Herbut1, Herbut2,abanin2006:nu0,brey2006:nu0}, Kharitonov~\cite{kharitonov2012:nu0} proposed a simple model that explains the experiments. In this model, in addition to the long-range Coulomb interaction, there are ``anisotropic" short-range interactions which have the symmetry $SU(2)_{\rm spin}{\otimes} U(1)_{\rm valley}$~\cite{alicea2006:gqhe,kharitonov2012:nu0}. Assuming only ultra-short-range interactions and using the Hartree-Fock (HF) approximation, Kharitonov found four phases~\cite{kharitonov2012:nu0}: (i) A spontaneous spin ferromagnet (FM) even in the absence of the Zeeman coupling $E_Z$, which does not break any of the lattice symmetries. (ii) A spin antiferromagnet that turns into a canted antiferromagnet (CAFM) upon including the Zeeman coupling $E_Z$. This state breaks the sublattice exchange symmetry. (iii) A bond-ordered phase (BO) that spontaneously breaks the $U(1)_{\rm valley}$ symmetry by filling states of both spins that superpose the valleys. This state breaks lattice translation symmetry. (iv) A charge density wave (CDW) state which is valley polarized. The CDW state also breaks the sublattice exchange symmetry. Thus, the phases either spontaneously break either the spin rotation symmetry or the lattice symmetry, but not both. The picture that emerges is that the system is in the CAF state at a purely perpendicular field, and undergoes a second-order phase transition into the FM phase at large $E_Z$.

Recently, spurred by the experimental tension between magnon scattering~\cite{Magnon_transport_Yacoby_2018, Spin_transport_Lau2018, Magnon_Transport_Assouline_2021, Magnon_transport_Zhou_2022} and scanning tunnelling microscopy (STM) results~\cite{li2019:stm, STM_Yazdani2021visualizing, STM_Coissard_2022, Yazdani_FQH_STM2023_1, Yazdani_FQH_STM2023_2}, anisotropic interactions that go beyond the ultra-short-range (USR) assumption of Kharitonov have been investigated~\cite{Das_Kaul_Murthy_2022, De_etal_Murthy2022, Stefanidis_Sodemann2023, Jincheng2024}. Such interactions are expected to be present on general grounds based on renormalization group (RG) thinking and have also been shown to be generated by Landau-level mixing in a recent perturbative calculation~\cite{Wei_Xu_Sodemann_Huang_LLM_SU4_breaking_MLG_2024}. Going beyond the USR assumption leads to new phases at $\nu{=}0$ that simultaneously break the lattice and spin-rotation symmetries, which offer a possible resolution to the experimental conundrums. Recent observations suggest that such LL-mixing-induced corrections to the anisotropic interactions also play an important role in other graphene systems~\cite{Khanna_BLG2023}.

A very convenient way to parameterize generic interactions projected to a particular LL manifold is through the Haldane pseudopotentials~\cite{Haldane_Pseudopot1983}. The Haldane pseudopotential $V_m$ measures the energy cost of a translation- and rotation-invariant two-body interaction in the relative angular momentum $m$ channel in a particular LL. A USR interaction corresponds to penalizing only the $m{=}0$ channel i.e., $V_{m}{=}V_{0}\delta_{m,0}$, whereas a generic interaction will have all $V_m{\neq}0$. 

Very recently, three of us presented phase diagrams for the fractional filling $\nu{=}{-}1/3$, going beyond USR interactions in a model where the anisotropic interactions had $V_0{\neq}0,~V_1{\neq}0$ and $V_{m{>}1}{=}0$~\cite{Jincheng2024}. Using a nontrivial extension of an earlier variational approach developed for USR interactions by Sodemann and MacDonald~\cite{Sodemann_MacDonald_2014}, using values of the anisotropic couplings natural for the ZLL, we found several phases that simultaneously break lattice and magnetic symmetries spontaneously. Some of these phases have no analog at $\nu{=}0$ and occur only for fractionally filled states. Additionally, several of these phases do not appear for USR couplings at $\nu{=}{-}1/3$, making it clear that the phase diagram for generic couplings is richer than for USR couplings. 

The goal of this work is two-fold. Firstly we extend our results with beyond-USR couplings in the ZLLs to fractional fillings near half-filling in higher LL manifolds, concentrating on the $n{=}1$ manifold. We consider filling factors with denominator 3 as well as 5. We also supplement our previous discussion in the ZLL manifold by considering fractions we had not previously considered. It should be noted at this point that the Hamiltonian with spin and valley Zeeman couplings, and anisotropic valley Ising valley $XY$ interactions are valid when projected to any LL manifold. However, there are several differences between the ZLL and the other manifolds. Even a bare USR interaction, when projected to a $n{\neq} 0$ manifold of LLs, will become effectively longer-ranged due to the single-particle wave functions of the higher manifolds. Also, because of the equal superpositions of the wave functions of either valley on the two sublattices in $n{\neq}0$ manifolds, sublattice symmetry breaking due to the substrate~\cite{KV_DGG_2013, KV_expt_2013, KV_HBN_Jung2015, KV_HBN_Jung2017} can be ignored to leading order. Last but not least, the values of the anisotropic interactions generated by projecting  USR anisotropic interactions into an $n{\neq}0$ LL-manifold are in a very different regime than the couplings obtained by projecting to the ZLLs. As we will explicitly show in a later section, the momentum-independent $q{=}0$ anisotropic interactions (the Hartree parts) vanish for $n{\neq}0$, while they are the largest ones for $n{=}0$. 

For some fractions, there are multiple ways to achieve the desired fractional filling. For $\nu{=}{-}3/5$, for example, we can fully fill one linear combination of LLs and fill an orthogonal linear combination with $2/5$. Such a two-component state (2CS) would be denoted ${\vec \nu}{=}\big(1,2/5,0,0\big)$. Alternatively, we could fill three orthogonal linear combinations with ${\vec \nu}{=}\big(1,[1/5,1/5],0\big)$, where the $[1/5,1/5]$ indicates that the two $1/5$ filled LLs are in an $SU(2)$ singlet. This would be a three-component state (3CS). 

In all, we obtain 15 different phases in the 2CS and 12 different phases in the 3CS. 

Here is a preview of our results: In the ZLLs there are no surprises for natural couplings, in the sense that the phases that are seen for $\nu{=}{-}2/3$ and $\nu{=}{-}3/5$ are the same as those seen in our earlier work~\cite{Jincheng2024}, with some quantitative differences in the phase diagrams. In Fig.~\ref{Fig: 1} we show the phase diagram for the 2CS at $\nu{=}{-}3/5$. Many new phases appear for natural couplings in the $n{=}1$ LL manifold that were not seen in the ZLLs. For example, the bond-ordered ferromagnet (BOFM), the bond-ordered canted antiferromagnet (BOCAFM), and the valley/spin antiferromagnet (V/SAF) occur only in the higher LL manifolds for natural couplings. 

The plan of the paper is as follows. In Sec.~\ref{sec: couplings_projections_to_LLs} we present the details of the symmetry-allowed USR couplings in graphene and the forms of the couplings that arise upon projection to the $n^{\rm th}$ LL-manifold. In Sec.~\ref{sec: spinor_ansatz} we will present our ansatz for the four linear combinations (spinors) which end up depending on seven angles. In this section, we also present our model for interactions beyond USR and the expression for the variational energy, which crucially needs some numerical input because the interactions are not USR. Sec.~\ref{sec: variational_wfns} presents the variational wave functions and the numerical results needed to complete the computation of the variational energy. In Sec.~\ref{sec: results_GLL1} we present our results for various fractions in the $n{=}1$ LL manifold. Finally, in Sec.~\ref{sec: discussion_summary} we end with a summary of our findings and some open questions. All the new results in the ZLLs and the comparison between phase diagrams in the ZLLs and the higher LL manifolds are relegated to appendices (see App.~\ref{app: A} and App.~\ref{app: B}) so as not to clutter the main manuscript.

\section{Couplings and their projections to Landau levels}
\label{sec: couplings_projections_to_LLs}

As mentioned above, the effective two-body interactions within any manifold of 4 degenerate LLs may always be written as non-USR valley XXZ terms~\cite{Das_Kaul_Murthy_2022, De_etal_Murthy2022, Stefanidis_Sodemann2023}. To treat the effective couplings in different manifolds (different $n$) on the same footing, in this work we consider a specific bare model of the microscopic interactions at $B_{\perp}{=}0$ and then project these to the LLs of interest. Note that corrections arising from LL mixing are not included in this bare model, leading to the conventional USR model within the $n{=}0$ manifold. Nevertheless, it does generate non-USR terms for the $|n|{>}0$ LLs due to the structure of the single-particle wave functions. We follow this approach for the $n\neq0$ LL manifolds. However, in the appendices, we will explicitly include non-USR interactions to obtain generic phase diagrams in the ZLLs and compare them to phase diagrams in the $n{\neq}0$ LL manifolds. In the following subsections, we first define the interaction model employed here, then provide the effective couplings resulting from the projection to the Landau level under consideration. 

\subsection{Short-range interactions at $B_{\perp}{=}0$}
\label{subsec: short_range_int}

Without a magnetic field, the linear Dirac spectrum of MLG makes short-range interactions irrelevant in the RG sense. 
The least irrelevant component of a generic (finite-ranged) interaction is the part that is independent of momentum (the local piece). 
In this work, we consider a model that includes all the local terms that are allowed by the symmetries of the system. 
For the case of MLG, a complete classification of such interactions was worked out in Refs.~\cite{RG1_Graphene_Aleiner_2007, RG2_Graphene_Aleiner_2008} which resulted in a model with 9 independent parameters. 

In the continuum limit of MLG, electrons carry three internal indices: $\sigma{=}\upa,\dwa$ for spin, $\tau{=}\bK,\bKp$ for the valley, and $\eta{=}A, B$ for the two sublattice sites. For brevity, we define a single 8-component field annihilation operator at position $\vecr$, $\Psi(\vecr)$, which includes all the internal flavors.  
Then a generic 2-body contact term has the form
\begin{align}
 \label{eq: Hint_B0_generic}
    H^{\rm generic}_{\rm int}= \frac{1}{2} \int d^2 \vecr :\Psi^{\dagger}({\vecr}) \mathcal{M}_{1} \Psi({\vecr}) \, 
  \Psi^{\dagger}({\vecr}) \mathcal{M}_{2} \Psi({\vecr}): \, ,
\end{align}
where $\mathcal{M}_{1,2}$ are $8{\times}8$ Hermitian matrices and $:$ $:$ denotes normal ordering. 
Most of the terms appearing in Eq.~\eqref{eq: Hint_B0_generic} are ruled out by the spatial and non-spatial symmetries relevant to MLG~\cite{RG1_Graphene_Aleiner_2007}. 
The former include translation by the lattice vectors, a 3-fold rotation about the $z$ (the out-of-plane) axis (forming the C${_3}$ group), and reflections about the $x$-$z$ and $y$-$z$ planes (forming two independent C$_{2}$ groups). 
The non-spatial symmetries include invariance under time-reversal operation as well as the spin $SU(2)$ transformations. 
Finally, the Fierz rearrangement identity allows the surviving terms to be expressed in a spin-independent form. 
Then the most general Hamiltonian describing the ultra-short-range interactions in MLG is
\begin{align} \label{eq: Hint_B0}
  H_{\rm int}= \frac{1}{2}\sum\limits_{ij} g_{ij} \int d^2 \vecr : \Big[ \Psi^{\dagger}({\vecr})\tau^i\eta^j \Psi({\vecr}) \Big]^{2} : \,\, .
\end{align} 
Here, $\tau$ and $\eta$ describe Pauli matrices acting in the space of valley and sublattice labels, and the sum over $i, j$ runs through $\{0, x, y, z\}$. 
We have employed the basis $\Psi^{\dagger} (\vecr) {=} \{ \Psi^{\dagger}_{\upa} (\vecr), \Psi^{\dagger}_{\dwa} (\vecr) \}$ with $\Psi^{\dagger}_{\sigma} (\vecr) {=} \{ \Psi^{\dagger}_{\bK A \sigma} (\vecr), \Psi^{\dagger}_{\bK B \sigma} (\vecr), \Psi^{\dagger}_{\bKp B \sigma} (\vecr), {-}\Psi^{\dagger}_{\bKp A \sigma} (\vecr) \}$, which makes the kinetic energy acquire a valley independent form. 
Note the ordering of the sublattice is opposite for the two valleys implying that terms that are diagonal in $\hat{\eta}$ couple the same (opposite) sublattice sites if they are diagonal (off-diagonal) in $\hat{\tau}$. 
The couplings $g$ in Eq.~\eqref{eq: Hint_B0} satisfy the following relations
\begin{align}
    &g_{x0} = g_{y0} \equiv g_{\perp 0}, \,\, 
    g_{xz} = g_{yz} \equiv g_{\perp z}, \label{eq: g1} \\
    &g_{0x} = g_{0y} \equiv g_{z \perp}, \,\,
    g_{zx} = g_{zy} \equiv g_{z \perp}, \label{eq: g2} \\
    &g_{xx} = g_{xy} = g_{yx} = g_{yy} \equiv g_{\perp \perp} \label{eq: g3} . 
\end{align}
Then Eq.~\eqref{eq: Hint_B0} is a sum of 16 terms, but there are only 9 independent couplings: $g_{00}$, $g_{0z}$, $g_{z0}$, $g_{zz}$ and the 5 defined in Eqs.~\eqref{eq: g1}-\eqref{eq: g3}. 
Each of these interactions respects the $U(1)$ valley symmetry which corresponds to the conservation of the difference of charge between $\bK$ and $\bKp$. 
This is the complete model of short-range interactions allowed in MLG~\cite{RG1_Graphene_Aleiner_2007}. 

However, only 6 of these 9 couplings are relevant for our analysis. 
First, $g_{00}$ is a completely diagonal term corresponding to a local density-density interaction. 
Hence, we may identify $g_{00}$ with the short-range component of the Coulomb interaction instead of treating it as an independent coupling. 
Furthermore, $g_{0z}$ and $g_{0 \perp}$ are diagonal in both the valley and spin index and therefore only contribute to the $SU(4)$ symmetric part of the interaction after projection to any manifold of LLs.   
Just like the Coulomb interaction, these two do not play any role in deciding the spin-valley ordering of the ground state. 
Henceforth in this paper, we shall ignore these three couplings and characterize the short-range interactions Eq.~\eqref{eq: Hint_B0} in terms of just 6 parameters, 
\begin{align} \label{eq: gij}
    g_{z0}, \, g_{zz}, \, g_{z\perp}, \, g_{\perp 0}, \, g_{\perp z}, \, g_{\perp \perp}.
\end{align}

These local couplings are likely sensitive to several effects, including the repulsive electronic interaction, electron-phonon couplings, and possibly some details of device configuration (such as the alignment to hBN, screening by gates, etc.). 
A recent work~\cite{Wei_Xu_Sodemann_Huang_LLM_SU4_breaking_MLG_2024} numerically estimated the values of $g_{ij}$, taking into account unscreened ($1/q$) Coulomb interactions and electron-phonon interactions, and found 2 of the 6 couplings ($g_{z 0}$ and $g_{\perp 0}$) to be vanishingly small. 
The other four couplings assume the following values~\cite{Wei_Xu_Sodemann_Huang_LLM_SU4_breaking_MLG_2024}, 
\begin{align}\label{g_values}
    &g_{zz} = 184 \text{ meV$\cdot$ nm}^{2} , \,\, 
    g_{z \perp} = -27 \text{ meV$\cdot$ nm}^{2}, \nn
    &g_{\perp z} = -26 \text{ meV$\cdot$ nm}^{2} , \,\, 
    g_{\perp \perp} = 269 \text{ meV$\cdot$ nm}^{2} . 
\end{align} 
To maintain generality, we shall not employ these specific values in our analysis. 
Nevertheless, we shall assume that $g_{z 0}, g_{\perp 0}$ (the 2 vanishing $g$'s) are much smaller in magnitude than the other 4 couplings. 
We shall also assume that the 4 large couplings are similar in magnitude, but not necessarily in sign, to the values reported in Ref.~\cite{Wei_Xu_Sodemann_Huang_LLM_SU4_breaking_MLG_2024}. 
We refer to the values of $g$ that satisfy these assumptions as ``natural'' couplings. 

\subsection{Model for interactions beyond ultra-short-range}
\label{sec: beyond USR}
Since we will always be dealing with interactions beyond USR in this work, we detour now to parameterizing such interactions. The Haldane pseudopotentials~\cite{Haldane_Pseudopot1983} offer a very compact way to characterize any translationally and rotationally invariant two-body interaction projected to a Landau level. The pseudopotential $V_m$ represents the strength of the interaction in the relative angular momentum channel $m$ of the two interacting electrons. An ultra-short-range interaction has only $V_0\neq 0$, with all other $V_m{=}0;~m{>}0$. In previous work by three of the present authors~\cite{Jincheng2024}, we used a model in the ZLLs with $V_0,~V_1{\neq} 0$, with all other $V_m{=}0;~m{>}1$. The four-fermion interaction matrix in the symmetric gauge, projected to a particular LL manifold, can be written as 
\begin{eqnarray}\label{Haldane_pseudopotentials}
&&{\hat V}=\sum\limits_{m_i} V_{m_1 m_2 m_3 m_4} c^{\dagger}_{m_1}c^{\dagger}_{m_2}c_{m_3}c_{m_4},\nonumber\\
&&V_{m_1 m_2 m_3 m_4}=\langle m_1, m_2\lvert \hat V\lvert m_3,m_4\rangle=\sum_m V_m{\mathbf U}^{(m)}_{m_1m_2m_3m_4},\nonumber\\
&&{\mathbf U}^{(m)}_{m_1m_2m_3m_4}=\sum_M \langle m_1,m_2|M,m\rangle\langle M,m|m_3,m_4\rangle,
\end{eqnarray}
As illustrative examples, the coefficients that combine two individual angular momenta $m_{1}$ and $m_{2}$ into the center-of-mass $M$ and relative angular $m$ momenta are 
\begin{eqnarray}
&\langle m,m'\lvert m+m',0\rangle=\frac{1}{\sqrt{2^{m+m'}}}\sqrt{\frac{(m+m')!}{m!m'!}}\nonumber\\
&\langle m,m'\lvert m+m'-1,1\rangle=\frac{m-m'}{\sqrt{2^{m+m'}}}\sqrt{\frac{(m+m'-1)!}{m!m'!}}.
\end{eqnarray}

\subsection{Effective Couplings After Projection}
\label{subsec: proj_coupling}

Our next task is to project the interactions in Eq.~\eqref{eq: Hint_B0} to the specific LL manifolds of MLG. 
For the $SU(4)$ symmetric Coulomb interaction, this projection was carried out in Refs.~\cite{Shibata09, Dora23}, in the context of MLG and also more generally for $J{-}$layer graphene. 
Here, we shall restrict our attention to the valley anisotropic short-range couplings in Eq.~\eqref{eq: gij}. 

The LLs of MLG have a non-trivial structure in the sublattice space and are best described in terms of a two-component spinor $|{\bf n}\rangle$. 
Here ${\bf n}$ corresponds to a given LL manifold including the sublattice indices. 
We shall define $n {=} |\bn|$ and use $|n\rangle$ to refer to the $n^{\rm th}$ scalar LL i.e., state of the non-relativistic LL with index $n$. 
Employing the basis $(A, B)$ for valley $\bK$ and $(B, {-}A)$ for $\bKp$, the spinor becomes valley independent and may be written as, 
\begin{align}
    &|{\bf n}=0\rangle = 
    \left( \begin{array}{c} |0\rangle \\ 0 \end{array} \right),\, 
    |{\bf n}\rangle= 
    \frac{1}{\sqrt{2}} \left( \begin{array}{c} |n\rangle \\ \text{sign}({\bf n}) |n-1\rangle \end{array} \right). 
    \label{eq: nLL wave functions}
\end{align}
The sublattice-valley locking of the ZLLs is manifest in the definition above (recalling the valley-dependent basis). 
On the other hand, the wavefunctions in the higher LLs have equal weight on both sublattices, albeit with a different orbital structure. 
We shall account for these sublattice features while writing the effective interactions in higher LLs by introducing appropriate form factors in the interaction. 

We begin by recalling that, in the symmetric gauge, the matrix elements of the density operator $\rho(\vecq){=}\sum_{j} e^{{-} i {\vecq} \cdot {\vec{r}_j}}$ in the scalar LLs may be written as,   
\begin{align} \label{eq: density_me}
 \langle n_{1}, m_{1} | e^{- i {\vecq} \cdot {\vecr}} | n_{2}, m_{2} \rangle = \rho_{m_{1} m_{2}} (\vecq) \rho_{n_{1} n_{2}} (\vecqs) ,
\end{align}
where $n_i$ are (scalar) Landau level indices, $m_i$ are symmetric gauge guiding center indices, and $\vecqs {\equiv} (q_{x}, {-}q_{y})$,  
the complex conjugate of $\vecq$ in the complex $q$-plane. 
For $m_{1} {\geq }m_{2}$ we have, 
\begin{align} 
    \rho_{m_{1} m_{2}} (\vecq) = \sqrt{\frac{m_{2}!}{m_{1}!}} \left(-i \frac{Q e^{i \theta_{\vecq}} }{\sqrt{2}} \right)^{m_{1} - m_{2}} e^{-\frac{Q^{2}}{4}} L_{m_{2}}^{m_{1} - m_{2}} \left( 
    \frac{Q^2}{2} \right)
\end{align}
where $Q{=}q\ell$ ($q{=}|\vecq|$), $\theta_{\vecq}$ is the polar angle of $\vecq$ and $\rho_{m_2 m_1}({\vecq}){=}\big(\rho_{m_1m_2}({-}{\vecq})\big)^*$. 
Note that these matrix elements satisfy $\rho_{n_{1} n_{2}} (\vecq) {=} \rho_{00}(\vecq) {\times} f_{n_{1} n_{2}}(\vecq)$ where (for $n_{1} {\geq} n_{2}$),
\begin{align} \label{eq: f_n1n2}
  f_{n_{1} n_{2}} (\vecq) = \sqrt{\frac{n_{2}!}{n_{1}!}} \left(-i \frac{Q e^{-i \theta_{\vecq}} }{\sqrt{2}} \right)^{n_{1} - n_{2}} L_{n_{2}}^{n_{1} - n_{2}} \left( 
    \frac{Q^2}{2} \right)
\end{align} 
As shown below, this property will enable us to map the electron dynamics in higher LLs, which have a sublattice-dependent orbital structure, to the $\bn{=} 0$ LL (which has a trivial sublattice structure) by absorbing an appropriate form factor in the interaction potential.

The local isospin density operator may be written in terms of the field operator ($\Psi$)  defined earlier as, 
\begin{align} \label{eq: rhoij_a}
\hat{\rho}_{ij} (\vecq) = \sum_{\sigma} 
  \int d^2 \vecr  \Psi_{\sigma}^{\dagger}({\vecr}) e^{-i \vec{q} \cdot \vec{r}} \tau^i\eta^j \Psi_{\sigma}({\vecr}). 
  \end{align}
The 4 components of $\Psi_{\sigma}$ may be labeled by $\{t, \xi\}$, where $t {=} 1,2$ corresponds to the $\bK, \bKp$ valleys and 
$\xi {=} 1,2$ refers to the two components of the spinors $|\bn\rangle$. 
Due to our choice of basis, $\xi {=} 1$ (or $2$) refers to different sublattice sites for different valleys ($t$). 
We further express $\Psi$ in the LL basis, defined in Eq.~\eqref{eq: nLL wave functions}, to get, 
$\Psi_{\xi t \sigma}({\vecr}) {=} \sum_{\bn,m} a_{\bn,\xi} \langle \vecr | \bn(\xi),m \rangle \cao_{\bn,m,t}$. 
Here, $\bn(\xi)$ is $n$ or $n{-}1$ for $\xi {=} 1$ or $2$, and $a$ is relative weight on each site: $a_{\bn {=} 0} {=} \delta_{\xi,1}$ and $a_{\bn {\neq} 0} {=} \frac{1}{\sqrt{2}} (\delta_{\xi,1} {+} \text{sign}(\bn) \delta_{\xi,2})$. 
After plugging this in Eq.~\eqref{eq: rhoij_a} and using Eq.~\eqref{eq: density_me}, the projection  of $\hat{\rho}_{ij} (\vecq)$ to $|\bn \rangle$ turns out to be, 
\begin{widetext}
\begin{align} 
  \hat{\rho}&_{ij}^{(\bn)} (\vecq) 
     = \sum_{\{m,t,\xi\}}
    \Big[\tau^{i}\Big]_{t_{1},t_{2}}  \Big[\eta^{j}\Big]_{\xi_{1},\xi_{2}}  \times
    a_{\bn, \xi_{1}} a_{\bn, \xi_{2}}  
    \rho_{m_{1}m_{2}}(\vecq)  \rho_{\bn(\xi_{1}),\bn(\xi_{2})}(\vecqs) \times 
      \cdo_{\bn,m_{1},t_{1},\sigma}\cao_{\bn,m_{2},t_{2},\sigma} \nonumber \\
     &= \bigg( \underbrace{\sum_{\{\xi\}} \Big[\eta^{j}\Big]_{\xi_{1},\xi_{2}} 
      a_{\bn, \xi_{1}} a_{\bn, \xi_{2}} 
     f_{\bn(\xi_{1}),\bn(\xi_{2})}(\vecqs)}_{F^{ }_{\bn,j} (\vecq)} \bigg) \times 
     \sum_{\{m\}} \bigg[ \rho_{m_{1}m_{2}}(\vecq) \rho_{00}(\vecqs) 
    \times \bigg( \underbrace{\sum_{\{t\}} \cdo_{\bn,m_{1},t_{1},\sigma} \Big[\tau^{i}\Big]_{t_{1},t_{2}}  \cao_{\bn,m_{2},t_{2},\sigma}}_{\hat{\tau}^{i}_{m_{1}m_{2}}}\bigg)\bigg]. \label{eq: rhoij}
\end{align} 
\end{widetext}
Here, we used Eq.~\eqref{eq: f_n1n2} in the second step and defined the form factor $F^{ }_{\bn,j} (\vecq)$ which completely accounts for the sublattice and orbital structure of the $\bn^{\text{th}}$ LL of MLG. Note that the matrix $\eta^{j}$ has been replaced by the scalar $F^{ }_{\bn,j}$.
For $\bn {=} 0$, the form factors are trivial: $F^{ }_{\bn{=}0,0} {=} F^{ }_{\bn{=}0,z} {=} 1$ and $F^{ }_{\bn{=}0,x} {=} F^{ }_{\bn{=}0,y} {=} 0$. 
In other cases, we have (using $Q {=} q\ell$), 
\begin{align}
  &F^{ }_{\bn \neq 0,0} = \frac{1}{2} \left[ L_{n} \left( 
    \frac{Q^2}{2} \right) + L_{n-1} \left( 
    \frac{Q^2}{2} \right) \right], \\
  &F^{ }_{\bn \neq 0,z} = \frac{1}{2} \left[ L_{n} \left( 
    \frac{Q^2}{2} \right) - L_{n-1} \left( 
    \frac{Q^2}{2} \right) \right], \\
  &F^{ }_{\bn \neq 0,x} = -i \, Q \cos(\theta_{\vecq}) \times \frac{\text{sign}(\bn)}{\sqrt{2 n}}  L^{1}_{n-1} \left( 
    \frac{Q^2}{2} \right) , \\
  &F^{ }_{\bn \neq 0,y} = -i \, Q \sin(\theta_{\vecq}) \times \frac{\text{sign}(\bn)}{\sqrt{2 n}}  L^{1}_{n-1} \left( 
    \frac{Q^2}{2} \right). 
\end{align}
The $j {=} 0$ result was previously derived in the case of Coulomb interactions~\cite{Dora23}. 
Note that the form factor is isotropic for $j {=} 0, z$, but not for $j {=} x, y$. 

We now write the contact anisotropic interactions in Eq.~\eqref{eq: Hint_B0} in terms of $\hat{\rho}_{ij}$ and generalize to a non-USR form, 
\begin{align} \label{eq: Han}
  H_{\rm an}= \frac{1}{2}\sum \limits_{ij} g_{ij} \int \frac{d^2 \vecq}{(2 \pi)^{2}} \, V_{ij} (q) 
  : \hat{\rho}_{ij} (\vecq) \hat{\rho}_{ij} (-\vecq) : \, , 
\end{align} 
where $V_{ij} (q)$ are generic isotropic interaction potentials, which reduce to a constant (1 in our convention) in the USR limit. 
These $V_{ij}$ are assumed to satisfy constraints analogous to those in Eqs.~\eqref{eq: g1}-\eqref{eq: g3} so that $H_{\rm an}$ respects the $U(1)$ valley symmetry of the system. 
After using Eq.~\eqref{eq: rhoij} to project $H_{\rm an}$ to $|\bn\rangle$, the projected Hamiltonian for the short-range anisotropic couplings in any LL manifold may be expressed in XXZ form as,    
\begin{align} 
  H^{(\bn)}_{\rm an} = \frac{1}{2}\sum \limits_{i} V^{(i,\bn)}_{m_1 m_2 m_3 m_4} \, :\hat\tau^i_{m_1m_4}\hat\tau^i_{m_2m_3}:\, .
  \label{eq: XXZform}
\end{align}
The constraints on $g_{ij}$ and $V_{ij}$ ensure that $V^{(x,\bn)} {=} V^{(y,\bn)} {\equiv} V^{(\perp,\bn)}$. 
Unlike Eq.~\eqref{eq: Han}, the projected Hamiltonian in Eq.~\eqref{eq: XXZform} involves only a sum over ($i$), the index of the valley matrices. 
The sublattice details have been incorporated in the matrix elements $V^{(i,\bn)}$ which are,
\begin{align} 
   \hspace{-1em}
   V^{(i,\bn)}_{m_1 m_2 m_3 m_4} = 
   \sum_{j} g_{ij}  
   \int &\frac{d^2 \vecq}{(2 \pi)^{2}} \, V_{ij}^{(\bn)} (\vecq) \times
   \nonumber \\ 
   &\rho_{m_{1}m_{4}}(\vecq) \rho_{m_{2}m_{3}}(-\vecq).  \label{eq: V1234}
\end{align}
Here, the effective potentials in the $\bn^{\text{th}}$ LL, $V_{ij}^{(\bn)}$, are,
\begin{align}
  V_{ij}^{(\bn)} (\vecq) = V_{ij} (q) e^{-\frac{1}{2} q^{2}\ell^{2}}
  \times F^{ }_{\bn,j} (\vecq) F^{ }_{\bn,j} (-\vecq).
\end{align} 
For $j {=} 0, z$, the effective interaction is clearly isotropic, 
\begin{align} 
\label{eq: V_i0}
  V_{i0}^{(\bn \neq 0)} (Q) = V_{i0} (Q) \frac{e^{-\frac{1}{2} Q^{2}}}{4}
  \left[ L_{n} \left( 
    \frac{Q^2}{2} \right) + L_{n-1} \left( 
    \frac{Q^2}{2} \right) \right]^{2}, \\ 
    \label{eq: V_iz}
      V_{iz}^{(\bn \neq 0)} (Q) = V_{iz} (Q) \frac{e^{-\frac{1}{2} Q^{2}}}{4}
  \left[ L_{n} \left( 
    \frac{Q^2}{2} \right) - L_{n-1} \left( 
    \frac{Q^2}{2} \right) \right]^{2}. 
\end{align}
For $j {=} x$ and $y$, $V_{ij}^{(\bn)}$ depends on $\cos^{2} (\theta_{\vecq})$ and $\sin^{2} (\theta_{\vecq})$ respectively. 
However, the matrix elements for a given $i$ in Eq.~\eqref{eq: V1234} depend only on a {\it scalar} sum over the sublattice index $j {=} \{0,x,y,z\}$. The symmetries of the system, which require $g_{ix} {=} g_{iy} {=} g_{i \perp}$ and $V_{ix} {=} V_{iy} {=} V_{i \perp}$, enable us to treat the sum of $j {=} x,y$ terms together, which is indeed circularly symmetric.
Thus we define, 
\begin{align} \label{eq: V_iperp}
  V_{i \perp}^{(\bn \neq 0)} &(Q) = V_{i \perp} (Q) e^{-\frac{1}{2} Q^{2}}
  \sum_{j = x,y} F^{ }_{\bn,j} (\vecq) F^{ }_{\bn,j} (-\vecq) \nonumber \\
  &= V_{i \perp} (Q) e^{-\frac{1}{2} Q^{2}}
  \times \frac{Q^{2}}{n} \left[ L^{1}_{n-1}  \left( \frac{Q^{2}}{2} \right) \right]^{2}. 
\end{align}  

Since the effective interaction potentials, defined in Eqs.~\eqref{eq: V_i0}-\eqref{eq: V_iperp}, have translation and rotational symmetries, we may characterize them in terms of Haldane pseudopotentials, as discussed in the previous section. 
Using Eq.~\eqref{Haldane_pseudopotentials} in Eq.~\eqref{eq: V1234}, we may write, 
\begin{align} \label{eq: um_A}
    V^{(i,\bn)}_{m_1 m_2 m_3 m_4} &= 
   \sum_m \bigg( \sum_{j} \frac{g_{ij}}{4 \pi \ell^{2}}  u^{(m)}_{ij,\bn} \bigg) {\bU}^{(m)}_{m_1m_2m_3m_4} \\ 
   &= 
   \sum_m u^{(m)}_{i,\bn} {\bU}^{(m)}_{m_1m_2m_3m_4}.  \label{eq: um_B}
\end{align}
In Eq.~\eqref{eq: um_A}, $j$ is summed over $\{ 0,z,\perp \}$, and we rescaled $g$ so that the pseudopotentials $u^{(m)}_{ij,\bn}$ are dimensionless. 
These potentials are easily found to be~\cite{Jain07}, 
\begin{align} \label{eq: um_C}
    u^{(m)}_{ij,\bn} = \int dQ \,  V_{ij}^{(\bn)}(Q) \times Q \, L_{m} (Q^{2}) e^{-Q^{2}}. 
\end{align}

The expressions above are valid in general. We now consider the special case of $V_{ij}(q) {=} 1$ implying that the microscopic interactions are of the USR form. Then the integral in Eq.~\eqref{eq: um_C} may be performed analytically given a value of $\bn$. 
Carrying out this procedure, we find that these pseudopotentials ($u^{(m)}_{ij,\bn}$) are finite for $m {\leq} 2n$ in the $\bn^{\text{th}}$ LL, implying that higher LLs support non-USR terms simply due to their single-particle wave functions. 
On the other hand, LL mixing is crucial to get non-USR interactions in the $\bn {=} 0$ manifold. In the following, we will assume that the couplings $V_{ij}$ are USR in the context of computing the phase diagrams for $\bn{\neq}0$ LL manifolds. Additionally, in the appendix, we treat some fractions in the ZLLs that were omitted in previous work by three of us~\cite{Jincheng2024}, and there we do assume non-USR interactions in the ZLLs, assumed to arise due to LL-mixing~\cite{RG_Sodemann_MacD2013, RG_Peterson_Nayak_2014, Wei_Xu_Sodemann_Huang_LLM_SU4_breaking_MLG_2024}.
Here are the couplings entering the XXZ Hamiltonian of Eq.~\eqref{eq: XXZform} broken up into pseudopotentials [as in  Eq.~\eqref{eq: um_B}] for $\bn {=} 0, 1$ and $2$, assuming purely USR $V_{ij}$ with $i{=}z,\perp$. 
\begin{align}
 &\quad u^{(0)}_{i,\bn = 0}= \frac{g_{i 0} + g_{i z}}{4 \pi \ell^2},&\label{n0_haldane}\\
 \nn
&\begin{cases}\label{n1_haldane}
   u^{(0)}_{i,\bn = 1} = \displaystyle\frac{5 g_{i 0} + g_{i z} + 4 g_{i \perp}}{32 \pi \ell^2}, \\[10pt]
    u^{(1)}_{i,\bn = 1} =\displaystyle \frac{g_{i 0} - g_{i z} - 2 g_{i \perp}}{16 \pi \ell^2}, \\[10pt] 
    u^{(2)}_{i,\bn = 1} = \displaystyle\frac{g_{i 0} + g_{i z}}{32 \pi \ell^2},
\end{cases}&\\
 \nn
&\begin{cases}\label{n2_haldane}
  u^{(0)}_{i,\bn = 2} =\displaystyle \frac{13 g_{i 0} + g_{i z} + 12 g_{i \perp}}{128 \pi \ell^2}, \\[10pt]
  u^{(1)}_{i,\bn = 2} = \displaystyle\frac{g_{i 0} - g_{i z} - 2 g_{i \perp}}{64 \pi \ell^2}, \\[10pt]
  u^{(2)}_{i,\bn = 2} =\displaystyle \frac{2 g_{i 0} + g_{i z} + g_{i \perp}}{32 \pi \ell^2}, \\[10pt]
  u^{(3)}_{i,\bn = 2} = \displaystyle\frac{3 g_{i 0} - 3 g_{i z} - 6 g_{i \perp}}{64 \pi \ell^2}, \\[10pt]
  u^{(4)}_{i,\bn = 2} = \displaystyle\frac{3 g_{i 0} + 3 g_{i z}}{128 \pi \ell^2}.
\end{cases}&
\end{align} 

The Haldane pseudopotentials presented above completely characterize the valley anisotropic interactions within a given LL manifold. 

\section{Spinor Ansatz, variational wave functions and energy}
\label{sec: spinor_ansatz}

\subsection{Spinor Ansatz, Order Parameters}
It is crucial to have an ansatz for the spinors that is capable of accessing all the phases. We will assume that the spinors can all be made real by separate $U(1)$ rotations in the spin and valley spaces. Since they are orthonormal, the first spinor can be described by three angles, the second spinor by two angles, the third by one angle, and the final one is fully determined. We thus need six angles to describe four real spinors. We use a generalization of the ansatz originally proposed to describe skyrmions~\cite{Doucot_Goerbig_Skyrmion_2008}, and subsequently used by many researchers~\cite{Lian_Rosch_Goerbig_2016, Lian_Goerbig_2017, Atteia_Goerbig_2021, Goerbig_nu0_Skyrmion_Zoo_2021, Hegde_Sodemann2022, Stefanidis_Sodemann2022, Stefanidis_Sodemann2023}. This generalization was developed in a previous work of three of the present authors~\cite{Jincheng2024}. We start with the following orthonormal set of spinors.
\bean\label{f_ansatz}
&&\lvert f_1\rangle =\cos\frac{\alpha_1}{2}\lvert\btau,\bs_a\rangle+\sin\frac{\alpha_1}{2}\lvert-\btau,-\bs_b\rangle,\nn
&&\lvert f_2\rangle =\cos\frac{\alpha_2}{2}\lvert\btau,-\bs_a\rangle+\sin\frac{\alpha_2}{2}\lvert-\btau,\bs_b\rangle,\nn
&&\lvert f_3\rangle =\sin\frac{\alpha_1}{2}\lvert\btau,\bs_a\rangle-\cos\frac{\alpha_1}{2}\lvert-\btau,-\bs_b\rangle,\nn
&&\lvert f_4\rangle =\sin\frac{\alpha_2}{2}\lvert\btau,-\bs_a\rangle-\cos\frac{\alpha_2}{2}\lvert-\btau,\bs_b\rangle\nn
&&\lvert\btau\rangle=\begin{pmatrix}
    \cos\frac{\theta_{\tau}}{2}\\
    e^{i\phi_{\tau}}\sin\frac{\theta_{\tau}}{2}
\end{pmatrix},\quad \quad 
  \lvert\bs\rangle=\begin{pmatrix}
    \cos\frac{\theta_{s}}{2}\\
    e^{i\phi_{s}}\sin\frac{\theta_{s}}{2}
\end{pmatrix},
\label{eq: First_ansatz}
\eean
where $|\btau,\bs\rangle {\equiv} |\btau\rangle{\otimes}|\bs\rangle$ with $\btau$ ($\bs$) as the vector in the valley (spin) Bloch sphere. We will always choose $\phi_{\tau},\phi_s{=}0,\pi$ since we have assumed the spinors to be real. This ansatz has five continuously varying angles, namely $\alpha_1,~\alpha_2,~\theta_{\tau},~\theta_a,~\theta_b$. From the counting argument given above, we know that we need at least six angles to describe all the spinors, so the above ansatz is not general enough. Therefore, we make a further set of rotations by introducing two more angles $\eta_1,~\eta_2$ as shown below.
\bean
&&|\tilde f_1\rangle=\cos\frac{\eta_1}{2}|f_1\rangle+\sin\frac{\eta_1}{2}|f_2\rangle,\nn
&&|\tilde f_2\rangle=-\sin\frac{\eta_1}{2}|f_1\rangle+\cos\frac{\eta_1}{2}|f_2\rangle,\nn
&&|\tilde f_3\rangle=\cos\frac{\eta_2}{2}|f_3\rangle+\sin\frac{\eta_2}{2}|f_4\rangle,\nn
&&|\tilde f_4\rangle=-\sin\frac{\eta_2}{2}|f_3\rangle+\cos\frac{\eta_2}{2}|f_4\rangle.
\label{eq: Second_ansatz}
\eean 
This description now uses seven continuously varying angles and is therefore slightly redundant, but is guaranteed to be complete. It should be noted that the same physical state, characterized by the projectors to the occupied states, can be described by multiple sets of angles due to the redundancy. However, we will characterize the states by the expectation values of various order parameters, which will not have this redundancy. \\
To differentiate all possible phases at filling $\vec \nu{=}(\nu_1,\nu_2,..,\nu_i,...)$, where $\nu_i$ is the filling of spinor $|f_i\rangle$, various order parameters $\hat O$'s are computed as 
\bean
\langle\hat O\rangle=\frac{\sum_i\nu_i {\rm Tr}(P_i\hat O)}{\sum_i\nu_i},
\eean 
where $P_i{=}| f_{i}\rangle\langle f_{i} |$ is the projector onto the spinors $|f_{i}\rangle$. The order parameter $\hat O$ will range over the following one-body operators, written schematically as $\sigma_z$ [representing ferromagnetic (FM) order], $\tau_z\sigma_x$ [representing canted antiferromagnetic (CAFM) order], $\tau_z$ [representing valley polarized (VP) order, which can also be interpreted as charge-density-wave (CDW) order in the ZLLs due to spin-valley locking], $\tau_x$ [representing intervalley coherence or bond order (BO), sometimes also termed Kekul\'e distorted (KD) order], $\tau_x\sigma_x$ [representing spin-valley entangled-X (SVEX) order], $\tau_y\sigma_y$ (SVEY), and $\tau_z\sigma_z$ [representing antiferromagnetic (AF) order].

\subsection{Hartree Fock Couplings in $\bn{=}1$ LL}

In this section, we will present a general algebraic expression for the variational energy. After projection to the $\bn{=}1$ manifold of LLs, with the spin degree of freedom taken into consideration, we can write the part of the Hamiltonian containing the anisotropic couplings as
\bean 
\hat H^{an}&=&\frac{1}{2}\left(\sum_{m=0}^2u_z^{(m)}{\mathbf U}^{(m)}_{m_1m_2m_3m_4}\right):\hat\tau^z_{m_1m_4}\hat\tau^z_{m_2m_3}:\nn
&&+\frac{1}{2}\left(\sum_{m=0}^2u_\perp^{(m)}{\mathbf U}^{(m)}_{m_1m_2m_3m_4}\right)\nn
&&\times :\hat\tau^x_{m_1m_4}\hat\tau^x_{m_2m_3}+\hat\tau^y_{m_1m_4}\hat\tau^y_{m_2m_3}:\nn
\hat\tau^i_{m_1m_4}&=&(\tau^i\otimes \sigma^0_{\text{spin}} )_{\alpha\beta}\hat c^\dagger_{m_1,\alpha}\hat c_{m_2,\beta},\nn
\alpha,\beta &=& K\uparrow,\ K\downarrow,\ K^\prime\uparrow,\ K^\prime\downarrow.
\eean 
To carry out the Hartree-Fock approximation we will need
\bean\label{S_m}
&&{\mathbf U}^{(0)}_{m_1m_2m_2m_1}={\mathbf U}^{(0)}_{m_1m_2m_1m_2},\nn
&&{\mathbf U}^{(1)}_{m_1m_2m_2m_1}=-{\mathbf U}^{(1)}_{m_1m_2m_1m_2},\nn
&&{\mathbf U}^{(2)}_{m_1m_2m_2m_1}={\mathbf U}^{(2)}_{m_1m_2m_1m_2},\nn
&&S^{(m)}=\frac{1}{N_\phi}\sum_{m_1m_2}{\mathbf U}^{(m)}_{m_1m_2m_2m_1}=2
\eean 

leading to the Hartree and Fock couplings being defined as 
\bean
u_{i,H}^{(m)}=(-1)^m u_{i,F}^{(m)}=u_{i}^{(m)}.
\eean 
To sum up, the overall Hartree and Fock couplings in the $\bn{=}1$ LL manifold are
\bean\label{usr_HF}
&&u_{\perp,H}=u^{(0)}_{\perp}+u^{(1)}_{\perp}+u^{(2)}_{\perp}=\frac{g_{\perp 0}}{4\pi\ell^2},\nn
&&u_{\perp,F}=u^{(0)}_{\perp}-u^{(1)}_{\perp}+u^{(2)}_{\perp}=\frac{g_{\perp 0}+2g_{\perp\perp}+g_{\perp z}}{8\pi\ell^2},\nn
&&u_{z,H}=u^{(0)}_{z}+u^{(1)}_{z}+u^{(2)}_{z}=\frac{g_{z0}}{4\pi\ell^2},\nn
&&u_{z,F}=u^{(0)}_{z}-u^{(1)}_{z}+u^{(2)}_{z}=\frac{g_{z0}+2g_{z\perp}+g_{z z}}{8\pi\ell^2}.
\label{eq: H and F XXZ couplings}\eean
We see that the Hartree parts of both types of anisotropic couplings depend only on $g_{z0},~g_{\perp0}$. The Hartree parts are defined as the values at $q{=}0$. From Eqs.~\eqref{eq: V_i0}, \eqref{eq: V_iz} and \eqref{eq: V_iperp} it can be seen that, for the $\bn{=}1$ manifold, at $q{=}0$ $V_{iz}(0){=}v_{i\perp}(0){=}0$, and only $V_{i0}{\neq} 0$. This has an important implication for natural couplings in the $\bn{=}1$ LL manifold. Recalling that $g_{i0}{=}0$ in the microscopic estimate~\cite{Wei_Xu_Sodemann_Huang_LLM_SU4_breaking_MLG_2024} we see that the Hartree parts of $u_i$ vanish in the $\bn{=}1$ LL manifold. From Eq.~\eqref{n2_haldane} one can easily verify that the Hartree parts of the two couplings depend only on $g_{i0}$. Based on Eqs.~\eqref{eq: V_i0}, \eqref{eq: V_iz} and \eqref{eq: V_iperp} it is straightforward to show that this holds for any $\bn{\neq}0$. Thus, there is no need to treat each LL manifold separately; one need only consider $\bn{=}0$ and one example for $\bn{\neq}0$. In this work, we focus on $\bn{=}1$ as the exemplar of all the $\bn{\neq}0$ manifolds.

\subsection{Variational Energy Functional}
The  FQHE variational states we consider are a product of integer and fractional parts in the Fock space representation, i.e.
\bean
|\Psi\rangle=|I\rangle\otimes|F\rangle=|I,F\rangle,\quad P_{I/F}=\sum_{i\in I/F}|f_i\rangle\langle f_i|.
\eean 
where $P_{I/F}$ are the density matrices constructed from spinors $|f_i\rangle$ occupied in the integer/fractional part of the state.
The variational energy we seek is the expectation value of the anisotropic interaction in this state. Since it is an expectation value, there are some simplifications. For example, one must annihilate and create particles in the same spinor and angular momentum state. Since our interaction is four-Fermi, there are only three possibilities: 
\begin{itemize}[left=0pt]
    \item (a) All four operators act on the integer-occupied spinors. It is clear that this part of the variational energy will have a Hartree-Fock form.
    \item (b) One pair of creation and annihilation operators acts on the integer-occupied spinors and the other acts on the fractionally occupied spinors.
    \item (c) All four operators act on fractionally occupied spinors.
\end{itemize}

   Therefore, the variational energy has three contributions. Schematically, 
\bean\label{energy_functional}
&&\quad\frac{1}{N_\phi}\langle\Psi|\hat H_{an}|\Psi\rangle=E_{II}+E_{IF}+
E_{FF},\nn
&&E_{II}\sim\langle I|\hat C^\dagger_I\hat C^\dagger_I\hat C_I\hat C_I|I\rangle,\ E_{IF}\sim\langle I|\hat C^\dagger_I\hat C^\dagger_F\hat C_I\hat C_F|F\rangle,\nn
&&E_{FF}\sim\langle F|\hat C^\dagger_F\hat C^\dagger_F\hat C_F\hat C_F|F\rangle,
\eean 
where $\hat C^\dagger_{I/F}/\hat C_{I/F}$ implicitly represent the creation/annihilation operators that will act on the integer/fractional occupied spinors.
As shown in previous work by three of us~\cite{Jincheng2024}, and reproduced for completeness in Appendix \ref{app: A}, it turns out that contributions (b) and (c) can also be expressed in a Hartree-Fock-like form. With this in mind, we define an expression representing a part of the variational energy depending on two subspaces defined by the projectors $P_1,\ P_2$ (which may be identical)

\bean\label{HF_form}
&&\mathcal{E}^{\rm an}(P_1,P_2,u_\perp,u_z)=\nn
&&\quad u_{\perp,H} \big[{\rm Tr}(P_1\tau_x){\rm Tr}(P_2\tau_x)+{\rm Tr}(P_1\tau_y){\rm Tr}(P_2\tau_y)\big]\nn
&&+u_{z,H}{\rm Tr}(P_1\tau_z){\rm Tr}(P_2\tau_z)\nn
&&-u_{\perp,F} \big[{\rm Tr}(P_1\tau_xP_2\tau_x)+{\rm Tr}(P_1\tau_yP_2\tau_y)\big]\nn
&&-u_{z,F}{\rm Tr}(P_1\tau_zP_2\tau_z),
\eean
then $E_{II}$ and $E_{IF}$ can be expressed as 
\bean\label{eii_eif}
&&E_{II}=\sum_m\frac{S^{(m)}}{2}\mathcal{E}^{\rm an}(P_I,P_I,u^{(m)}_\perp,u^{(m)}_z)\nn
&&E_{IF}=\sum_m\frac{S^{(m)}}{2}\nn
&&\quad\quad\times\begin{cases}
   2\times\nu \mathcal{E}^{\rm an}(P_I,P_F,u^{(m)}_\perp,u^{(m)}_z),\ |F\rangle=|\nu\rangle, \\
   \\
    2\times\frac{\nu}{2} \mathcal{E}^{\rm an}(P_I,P_F,u^{(m)}_\perp,u^{(m)}_z),\ |F\rangle=|\frac{\nu}{2},\frac{\nu}{2}\rangle,
\end{cases}\nn
\eean 
where $S_m$ is defined in Eq.\eqref{S_m}. The form of $E_{IF}$ was initially derived in Ref.~\cite{Sodemann_MacDonald_2014} for USR interactions.
The evaluation of $E_{FF}$, first obtained in Ref.~\cite{Jincheng2024}, is more involved, dependent on the specific properties of the FQHE states, and requires a numerical calculation. After presenting the details of the FQH wave functions we use in the next section, we will explicitly write out $E_{FF}$ as a Hartree Fock form of Eq.~\eqref{HF_form} for the Haldane pseudopotentials considered in this work, which will be justified in Appendix \ref{app: A}.\\
Besides the anisotropic interaction, we will also include a Zeeman energy in the Hamiltonian, 
\bean
\hat H_z=-E_z\hat\sigma^z,\quad \hat\sigma^z=(\sigma^z)_{\alpha\beta}\hat c^\dagger_{m,\alpha}\hat c_{m,\beta}.
\eean 
In the ZLLs, due to valley-sublattice locking, there is a valley Zeeman term as well, arising from the sublattice potential induced by the alignment with the encapsulating HBN~\cite{KV_DGG_2013, KV_expt_2013, KV_HBN_Jung2015, KV_HBN_Jung2017}. Such a term does not occur to leading order for the $\bn\neq0$ LL manifolds. Thus we will ignore such a term in what follows. 
Throughout this work, we choose to fix the magnetic field at $B_{\perp}{=}10~{\rm Tesla}$, implying a Zeeman energy of $E_z{=}(1/2)g_{B}\mu_B B_{\perp}{=}0.58~{\rm meV}$ [the $g$-factor $g_{B}{=}2$ for graphene]. In later phase diagrams, we will take the $E_z$ as the energy unit, i.e. $E_z{=}1$.

\section{FQHE Variational wave functions}
\label{sec: variational_wfns}
This section discusses the variational wave functions we deploy to capture the fractionally filled states. We will use trial states constructed from the composite fermion (CF) theory~\cite{Jain89} since previously it has been shown that CF states give a good representation of the Coulomb ground state in the LLs of our interest~\cite{Dev92a, Wu93, Jain07, Shibata09, Balram15c, Kim19, Balram21b}.

The CF theory postulates that the FQHE of electrons can be understood as the IQHE of CFs which are bound states of an electron and even number ($2p$) of vortices~\cite{Jain89}. The FQHE state at $\nu{=}n/(2pn{\pm }1)$ is mapped to the $\nu^{\rm CF}{=}n$ state of CFs carrying $2p$ vortices. In the presence of additional degrees of freedom, such as spin, valley, sub-lattice, etc., $n{=}\sum_{\lambda}n_{\lambda}$ where $\lambda$ indexes the different components and $n_{\lambda}$ is the number of filled CF-LLs of component $\lambda$. Restricting to the case of two components and using the notation for spin, we have $n{=}n_{\uparrow}{+}n_{\downarrow}$. We emphasize that ``spin" here is not the actual electron spin, but merely a schematic label for the two orthonormal spinors participating in the FQH state in the actual problem of interest. This state, denoted as $[n_{\uparrow}, n_{\downarrow}]_{{\pm }2p}$~\cite{Balram15}, is an eigenstate of the total spin $\vec{S}$ and carries a polarization $\gamma{=}(n_{\uparrow}{-}n_{\downarrow})/(n_{\uparrow}{+}n_{\downarrow}){\equiv}S_{z}/(N/2)$ ($N$ is the total number of electrons), where without loss of generality, we can assume that $n_{\uparrow}{\geq}n_{\downarrow}$.

The electronic ground state at $\nu{=}n/(2pn{\pm }1)$ is described by Jain's CF wave function~\cite{Jain89}
\begin{equation}
 \Psi^{{\rm Jain}[n_{\uparrow}, n_{\downarrow}]_{\pm 2p}}_{n/(2pn\pm 1)}=\mathcal{P}_{\rm LLL}\Phi_{{\pm}n_\uparrow}\Phi_{{\pm}n_\downarrow}\prod_{1\leq j<k \leq N}(Z_j{-}Z_k)^{2p}, 
 \label{eq: Jain_wfn}
\end{equation}
where $\{Z\}{=}\{z^{\uparrow}\}{\cup}\{z^{\downarrow}\}$ denotes the set of \emph{all} $N$ electronic coordinates. The complex coordinate $z^{\sigma}_j$ parameterizes the position of the $j$th electron with spin $\sigma$, $\Phi_{n}$ is the Slater determinant wave function of $n$ filled LLs with $\Phi_{{-}|n|}{=}[\Phi_{|n|}]^*$, and $\mathcal{P}_{\rm LLL}$ is an operator that implements projection to the LLL. In particular, the wave function of one filled LL of up spins is given by the Laughlin-Jastrow factor $\Phi_{n_\uparrow{=}1}{=}\prod_{1{\leq}j{<}k{\leq}N_{\uparrow}}(z^{\uparrow}_j{-}z^{\uparrow}_k)$, where $N_{\uparrow}$ is the number of up-spin electrons. The state $\Phi_{n}{\equiv}\Phi_{{\pm}n_\uparrow}\Phi_{{\pm}n_\downarrow}$ is the IQHE state and the Laughlin-Jastrow factor $\prod_{1\leq j<k \leq N}(Z_j{-}Z_k)^{2p}$ does the vortex-attachment to turn electrons into CFs.  

When $n_{\uparrow}{=}n_{\downarrow}{=}1$ and we take the ${+}$ sign in Eq.~\eqref{eq: Jain_wfn}, the Jain wave functions reduce to the Halperin $(2p{+}1,2p{+}1,2p)$ wave functions at $\nu{=}2/(4p{+}1)$~\cite{Halperin83} which are conventionally written as
\begin{eqnarray}
 \Psi^{{\rm Halperin}(2p+1,2p+1,2p)}_{2/(4p+1)}&=&\prod_{j<k}(z^{\uparrow}_j{-}z^{\uparrow}_k)^{2p+1}
 \prod_{j<k}(z^{\uparrow}_j{-}z^{\uparrow}_k)^{2p+1} \nonumber \\ 
 &&\times \prod_{j,k}(z^{\uparrow}_j{-}z^{\downarrow}_k)^{2p}.
 \label{eq: Halperin_wfn_2p1_2p1_2p}
\end{eqnarray}
Of particular interest to us is the Halperin $(3,3,2)$ spin-singlet state at $\nu{=}2/5$ which can be generated by diagonalizing its model-Hamiltonian $V_{m}{=}\delta_{m,0}{+}\delta_{m,1}$~\cite{Yoshioka89}. Except for the Halperin $(3,3,2)$ state, we use the ground states obtained from the exact diagonalization of the ideal Coulomb interaction in the LLL to generate the CF states since for all accessible systems it has been shown previously that the CF wave functions provide a near-perfect representation of the LLL Coulomb ground state~\cite{Ambrumenil88, Dev92a, Wu93, Jain07, Balram16b, Yang19a, Balram20b, Balram21b}.  

The information we need on these wave functions is the average pair amplitude of electrons in them in various relative angular momentum states. To get that, we compute these average pair amplitudes for many finite systems on the Haldane sphere~\cite{Haldane_Pseudopot1983} by evaluating expectation values of the relevant pseudopotential Hamiltonian in Fock-space and extrapolating those results to the thermodynamic limit. We note a couple of special cases here: (i) the $[1,0]_{2}$ 1/3 Laughlin state~\cite{Laughlin_1983} has $\langle\mathcal{P}_{m{\leq}2}\rangle{=}0$ while the $[1,0]_{4}$ 1/5 Laughlin state has $\langle\mathcal{P}_{m{\leq}4}\rangle{=}0$~\cite{Trugman85}, where $\langle\mathcal{P}_{m}\rangle$ is the average pair amplitude in relative angular momentum $m$, and (ii) the $[1,1]_{2}$ 2/5 Halperin $(3,3,2)$ spin-singlet state has $\langle\mathcal{P}_{m{\leq}1}\rangle{=}0$ and by this we mean $\langle \mathcal{P}^{\sigma, \sigma^{\prime}}_{m{\leq}1} \rangle{=}0$ for all values of $\sigma$ and $\sigma^{\prime}$, where $\mathcal{P}^{\sigma, \sigma^{\prime}}_{m}$ is the pair amplitude for two electrons in relative angular momentum $m$, one with spin $\sigma$ and the other with spin $\sigma^{\prime}$. All the relevant nonzero average pair amplitudes are given in Tables~\ref{ED_polar} and \ref{ED_singlet} (For 2/3, these numbers were already presented in Ref.~\cite{Jincheng2024}.). At 2/3 and 2/5, in the $n{=}0$ LL, the singlet state has a lower bare Coulomb energy than the fully polarized state~\cite{Balram15a} while in the $n{=}1$ LL the fully polarized state has lower Coulomb energy than the singlet~\cite{Balram15c}. Throughout this work, we do not consider any decoupled states, for example, the direct product of two 1/3 Laughlin states at 2/3 or the direct product of two 1/5 Laughlin states at 2/5, since these have a much higher Coulomb energy than the corresponding fully polarized and singlet states~\cite{Faugno20}.

\subsection{Expression for $E_{FF}$ for a single fractional component  $|F\rangle=|\nu\rangle$}

When the FQHE state is characterized by the fractional occupation of a single spinor, the form of $E_{FF}$, as shown in Appendix~\ref{app: A}, is 
\bean 
E_{FF}=\frac{1}{2}\sum_m\frac{\langle\nu|\hat {\mathbf U}^{(m)}|\nu\rangle}{N_\phi}\frac{1}{2}
\mathcal{E}^{\rm an}(P_F,P_F,u^{(m)}_\perp,u^{(m)}_z),\nn
\eean 
where $\hat {\mathbf U}^{(m)}{=} {\mathbf U}^{(m)}_{m_1m_2m_3m_4}\hat c^\dagger_{m_1}\hat c^\dagger_{m_2}\hat c_{m_3}\hat c_{m_4}$, and the spinor index has been omitted because only one fractionally filled spinor appears. Furthermore, the projector $P_F$ projects to this spinor. 
\begin{table}[h]
\centering
\begin{tabular}{|p{1cm}|p{3cm}|p{3cm}|}
        \hline
         $m$ & $\langle \frac{2}{3}|\hat {\mathbf U}^{(m)}|\frac{2}{3}\rangle/N_\phi$ & $\langle \frac{2}{5}|\hat {\mathbf U}^{(m)}|\frac{2}{5}\rangle/N_\phi$\\
        \hline
        0 & 0 & 0\\
        \hline
        1 & 4/3~\cite{Jincheng2024} & 0.1008\\
        \hline
        2 & 0&0 \\
        \hline             
\end{tabular}
\caption{Average pair amplitudes of the spinor-polarized $2/3$ and $2/5$ FQHE states for relative angular momentum $m{=}0,~1,~2$. }\label{ED_polar}
\end{table}\\
For one-component fractional states, the pair amplitudes in even relative angular momenta are zero due to fermionic antisymmetry.

\subsection{Expression for $E_{FF}$ for a two-component fractional singlet state, $|F\rangle=|[\frac{\nu}{2},\frac{\nu}{2}]\rangle$}
As we will show in Appendix~\ref{app: A}, we can express $E_{FF}$ for a singlet FQHE state as
\bean\label{Eff_polarized}
E_{FF}=\frac{1}{2}&&\sum_m\frac{\langle  \hat V^{(m)}\rangle}{[2-(-1)^m]N_\phi}\nn
&&\times\frac{1}{2}
\mathcal{E}^{\rm an}(P_F,P_F,u^{(m)}_\perp,u^{(m)}_z).
\eean 
Here the projector $P_F$ is the sum of the projectors of the two states participating in the fractional singlet. Note that $P_F$ is invariant under arbitrary $SU(2)$ rotations within the fractional subspace, implying that $E_{FF}$ is also invariant under $SU(2)$ transformations. Since such transformations do not change the total state, this is a necessary condition. 
\begin{table}[h]
\centering
\begin{tabular}{|p{1cm}|p{3.5cm}|p{3.5cm}|}
        \hline
         $m$ & $\langle [\frac{1}{3},\frac{1}{3}]|\hat V^{(m)}|[\frac{1}{3},\frac{1}{3}]\rangle/N_\phi$ & $\langle [\frac{1}{5},\frac{1}{5}]|\hat V^{(m)}|[\frac{1}{5},\frac{1}{5}]\rangle/N_\phi$\\
        \hline
       0 & 0~\cite{Jincheng2024} & 0\\
        \hline
        1 & 0.9827~\cite{Jincheng2024} & 0\\
        \hline
        2 & 0.976& 0.433\\
        \hline            
\end{tabular}
\caption{Average pair amplitudes of the spinor-singlet $2/3$ and $2/5$ FQHE states for relative angular momentum $m{=}0,~1,~2$. }\label{ED_singlet}
\end{table}

\section{Results for the $\bn{=}1$ Landau level}
\label{sec: results_GLL1}
In the graphene $\bn{=}1$ LL manifold, assuming USR interactions $V_{ij}$, there are 3 nonzero Haldane pseudopotentials. Eq.~\eqref{n1_haldane} relates the Haldane pseudopotentials $u_i^{(m)}$ to the six independent bare couplings $g_{ij}$'s, which have recently been microscopically estimated~\cite{Wei_Xu_Sodemann_Huang_LLM_SU4_breaking_MLG_2024}. In this estimate, it transpires that $g_{i0}{=}0$. We will stay ``close'' to this estimate by considering values of $g_{i0}$ which are at least an order of magnitude smaller than the other couplings. Operationally, we will fix $g_{z0},~g_{\perp0}$ at small values, which leaves four other couplings to vary freely. The coupling constant space is too large to visualize, so we will fix two of the couplings, say $g_{zz},~g_{\perp z}$ (or $g_{z\perp},~g_{\perp\perp}$), and plot two-dimensional sections of this four-dimensional space by varying $g_{z\perp}$ and $g_{\perp\perp}$ (or $g_{zz},~g_{\perp z}$). Note that this is physically equivalent to varying the Fock part of the XXZ couplings $u_z,~u_{\perp}$

\subsection{Phases diagrams for two-component states $\vec\nu{=}(1,\nu)$}

In Figs.~\ref{Fig: 1} and \ref{Fig: 3}, we present the phase diagrams at filling $\vec \nu{=}(1,2/3),\ (1,2//5)$, where $g_{\perp 0}$ and $g_{z0}$ are fixed and small, of the order of ${\sim} 10\text{meV}{\cdot }\text{nm}^{2}$ with both positive and negative signs. Besides the new phases found in recent work by three of us~\cite{Jincheng2024}, here we also obtain several additional phases. This may seem a bit strange for the following reason: The XXZ Hamiltonian [Eq.~\eqref{eq: XXZform}] describes the Hamiltonian of the anisotropic couplings in any LL manifold. Thus, all phases should appear in any given LL manifold. Why is it that we did not see the additional phases in our recent work~\cite{Jincheng2024}? The resolution of this seeming contradiction is contained in the notion of ``natural" couplings in a given LL manifold. An examination of the expressions for the Hartree XXZ couplings in Eq.~\eqref{eq: XXZform} shows that for $\bn{\neq} 0$ these couplings vanish to leading order because $g_{z0},~g_{\perp0}$ vanish at leading order~\cite{Wei_Xu_Sodemann_Huang_LLM_SU4_breaking_MLG_2024}. In contrast, in the ZLLs, the Hartree couplings are expected to be large. In previous work by the three of us~\cite{Jincheng2024}, we focused on the regime of natural couplings in the ZLLs, whereas here we examine natural couplings in higher LL manifolds.

In total, we have 15 different phases for the 2CSs. We organize the phases into three classes. In the first class, the occupied spinors can be written as direct products $|\btau\rangle{\otimes}|\bs\rangle$, where $\btau{=}\pm{\hat e}_z{=}{\bf K}/{\bf K}'$. The spinors are thus valley polarized (VP) and we will prefix the corresponding phases by VP, or sometimes V. Note that in the ZLLs we had the correspondence ${\rm VP}{\equiv}{\rm CDW}$ due to valley-sublattice locking, but this is no longer the case for $\bn{\neq} 0$ LL manifolds. In the second class, the spinors can also be written as direct products but now $\btau$ lies in the equatorial plane (or close to it), signifying intervalley coherence, which always implies some form of lattice symmetry breaking because the coherence is between two momenta not separated by a reciprocal lattice vector. Kekul\'e order, or more generally bond order, are specific instances of lattice symmetry breaking, which are determined by nonlinear terms in the Landau-Ginzburg theory. To keep the notation as simple as possible, we will use bond order (BO) as a shorthand for any lattice symmetry breaking. The phases with intervalley coherence are prefixed by BO. The third class has occupied spinors that cannot be written as direct products. We call such phases spin-valley entangled (SVE). With this explanation, the phases we find are listed below, along with the occupied spinors. The first spinor in each phase is fully occupied, while the second is partially occupied. 

The spinors of each phase are listed below,
\begin{itemize}[left=0pt]
    \item Valley Polarized Ferromagnet (VPFM): $|K,\uparrow\rangle,\  |K^\prime,\uparrow\rangle$,
    \item Valley Polarized Antiferromagnet (VPAFM): \\$|K,\uparrow\rangle,\ |K,\downarrow\rangle$
    \item Bond Order (BO): $|\hat e_x,\uparrow\rangle,\ |\hat e_x,\downarrow\rangle$
    \item Bond-Ordered Ferromagnet (BOFM): $|\hat e_x,\uparrow\rangle,\  |-\hat e_x,\uparrow\rangle$,
    \item Valley/Spin Antiferromagnet (V/SAFM): $|K,\uparrow\rangle,\ |K^\prime,\downarrow\rangle$
    \item Bond-Ordered AFM (BOAFM): $|\hat e_x,\uparrow\rangle,\ |-\hat e_x,\downarrow\rangle$
    \item Canted Antiferromagnet (CAFM): $|K,\nwarrow_1\rangle,\ |K^\prime,\nearrow_2\rangle$
    \item Bond-ordered CAFM (BOCAFM): \\$|\hat e_x,\nwarrow_1\rangle,\ |-\hat e_x,\nearrow_2\rangle$
    \item Canted Bond Order (CBO): $|\nwarrow_1,\uparrow\rangle,\ |\nearrow_2,\downarrow\rangle$ 
    \item SVE1:
    $\cos{\frac{\alpha}{2}}|K^\prime,\uparrow\rangle-\sin{\frac{\alpha}{2}}|K,\downarrow\rangle,\ |K,\uparrow\rangle$
    \item SVE2: There is no simple description for this phase, all order parameters are nonzero.
    \item SVE3: $\cos\frac{\alpha}{2}|K,\uparrow\rangle-\sin\frac{\alpha}{2}|K^\prime,\downarrow\rangle,\ |K,\downarrow\rangle,$
    \item SVE4:\\
    $ \cos\frac{\alpha_1}{2}|\hat e_x,\uparrow\rangle-\sin\frac{\alpha_1}{2}|-\hat e_x,\downarrow\rangle$,\\ $\cos\frac{\alpha_2}{2}|\hat e_x,\downarrow\rangle-\sin\frac{\alpha_2}{2}|-\hat e_x,\uparrow\rangle\quad$
    \item SVE5:\\
    $\cos\frac{\alpha_1}{2}|K,\uparrow\rangle-\sin\frac{\alpha_1}{2}|K^\prime,\downarrow\rangle$, $\sin\frac{\alpha_1}{2}|K,\uparrow\rangle+\cos\frac{\alpha_1}{2}|K^\prime,\downarrow\rangle$
    \item SVE6: \\
    $\cos\frac{\alpha_1}{2}|K,\downarrow\rangle-\sin\frac{\alpha_1}{2}|K^\prime,\uparrow\rangle$, $\cos\frac{\alpha_2}{2}|K,\uparrow\rangle-\sin\frac{\alpha_2}{2}|K^\prime,\downarrow\rangle$.   
\end{itemize}
In the spinor expression above when a canted spin is shown we identify
\bean
|\nwarrow_1\rangle=\begin{pmatrix}
    \cos\frac{\theta_1}{2}\\
    \sin\frac{\theta_1}{2}
\end{pmatrix},\quad
|\nearrow_2\rangle=\begin{pmatrix}
    \cos\frac{\theta_2}{2}\\
    -\sin\frac{\theta_2}{2}
\end{pmatrix}.
\eean 
All nontrivial angles are determined by the minimization of the energy functional. Let us now turn to the phase diagrams.

Below we will present a set of phase diagrams for the fractions $(1,\frac{2}{5},0,0)$ and $(1,\frac{2}{3},0,0)$. As mentioned before, the nominal values of $g_{z0},~g_{\perp0}$ are zero in the microscopic estimate~\cite{Wei_Xu_Sodemann_Huang_LLM_SU4_breaking_MLG_2024}. However, certain phases become nearly degenerate when we set $g_{z0}{=}g_{\perp0}{=}0$. To discriminate between these phases we assume that $g_{z0},~g_{\perp0}$ are nonzero, but much smaller than the other couplings in magnitude. We will generate phase diagrams for all four possible signs of $g_{z0},~g_{\perp0}$. 

We also have another choice to make. There are four other couplings $g_{zz},~g_{\perp z},~g_{z\perp},~g_{\perp\perp}$ that can be varied. Since a four-dimensional coupling constant space is very difficult to visualize, we take two-dimensional sections. We use the microscopic estimate~\cite{Wei_Xu_Sodemann_Huang_LLM_SU4_breaking_MLG_2024} to fix values for two of these four couplings and vary the other two. It is important to realize that the XXZ Hamiltonian projected to a LL-manifold has only two independent types of couplings $u_z,~u_\perp$, each coming with its array of Haldane pseudopotentials. Examining Eq.~\eqref{n1_haldane}, which we reproduce here for convenience,
\begin{eqnarray}\label{eq: Hspeudo_n=1}
&&u^{(0)}_{\perp} =\frac{5 g_{\perp 0} + g_{\perp z} + 4 g_{\perp \perp}}{32 \pi \ell^2},\ 
   u^{(0)}_{z} = \frac{5 g_{z 0} + g_{z z} + 4 g_{z \perp}}{32 \pi \ell^2},\nn
&&u^{(1)}_{\perp} = \frac{g_{\perp 0} - g_{\perp z} - 2 g_{\perp \perp}}{16 \pi \ell^2},\ 
    u^{(1)}_{z} = \frac{g_{z 0} - g_{z z} - 2 g_{z \perp}}{16 \pi \ell^2},\nn
&&u^{(2)}_{\perp} =\frac{g_{\perp 0} + g_{\perp z}}{32 \pi \ell^2},\ 
    u^{(2)}_{z} = \frac{g_{z 0} + g_{z z}}{32 \pi \ell^2},
\end{eqnarray}
we see that the $m{=}2$ Haldane pseudopotentials $u^{(2)}_i$ depend only on $g_{i0}$ and $g_{iz}$. We will present two types of sections through the 4-dimensional space of $g_{zz},~g_{\perp z},~g_{z\perp},~g_{\perp\perp}$. In the first type of section (type I), we will keep $g_{\perp z},~g_{zz}$ fixed and plot phase diagrams in $g_{\perp\perp},~g_{z \perp}$ space. As can be seen from the above equation, this keeps the $m{=}2$ pseudopotentials fixed while varying the $m{=}0,1$ pseudopotentials. In the second type of section (type II), we keep $g_{\perp\perp},~g_{z\perp }$ fixed while plotting the phase diagram in the space of $g_{\perp z},~g_{zz}$.  This will allow the $m{=}2$ pseudopotentials to vary. 

For 2CSs such as $(1,\frac{2}{5},0,0)$ or $(1,\frac{2}{3},0,0)$, because there is only one fractionally occupied spinor, the even angular momentum Haldane pseudopotentials do not participate in the $E_{FF}$ part of the variational energy. Thus, we expect the two types of sections to produce very similar phase diagrams. However, for the 3CSs where two fractionally occupied spinors occur, the even $m$ Haldane pseudopotentials do affect the variational energy, and therefore we expect the two types of sections to produce qualitatively different phase diagrams.

With these preparatory remarks, we are now ready to present phase diagrams. We will present the phase diagrams for $(1,\frac{2}{5},0,0)$ before the phase diagrams for $(1,\frac{2}{3},0,0)$ because more phases appear in the former. In Fig.~\ref{Fig: 1} we take a type I section [fixing $g_{zz},~g_{\perp z}$ at the values of the micropscopic estimate~\cite{Wei_Xu_Sodemann_Huang_LLM_SU4_breaking_MLG_2024}, Eq.~\eqref{g_values}] and plot the phase diagram in the $g_{z\perp},~g_{\perp\perp}$ plane. As mentioned before, to break degeneracies, we need to assign small values to $g_{z0},g_{\perp0}$. Panel (a) of Fig.~\ref{Fig: 1} has $g_{z0},g_{\perp0}{<}0$, panel (b) has $g_{z0}{>}0,g_{\perp0}{<}0$, panel (c) has $g_{z0}{<}0,g_{\perp0}{>}0$, and finally panel (d) has $g_{z0},g_{\perp0}>0$. The VPFM, BOFM, CAF, BOCAF, and the SVE1, SVE2, and SVE4 phases are present in all four cases, albeit with slightly different regions of stability. We find that $g_{\perp0}{<}0$ stabilizes the VPAF phase in the second quadrant of the $g_{z\perp},~g_{\perp\perp}$ plane, while $g_{\perp0}{>}0$ stabilizes the V/SAF phase in the same region. Similarly, $g_{z0}{<}0$ stabilizes the BO phase in the third/fourth quadrant, while $g_{z0}{>}0$ stabilizes the BOAF in the same region. We also note that $g_{\perp0}$ seems necessary to stabilize the SVE5 phase. The SVE6 phase does not appear in this particular type of section. 

In the lower four panels of Fig.~\ref{Fig: 1}, we plot the order parameters vs $g_{\perp\perp}$ at fixed $g_{z\perp}$ for the phase diagrams corresponding to the top four panels. In Fig.~\ref{Fig: 1}(e) (corresponding to the phase diagram of Fig.~\ref{Fig: 1}(a) along the black horizontal line shown there) the system starts in the VPAF  phase (red), makes a second-order transition to an extremely narrow region of the CBO phase, then another second-order transition into the BO phase (green). Next, it undergoes a first-order transition into the SVE2 phase (orange), and thence into a narrow region of the SVE4 phase (pink) via a second-order transition, and finally enters the BOFM phase via another second-order transition. The only first-order transition is between the simple phase BO and SVE2, which is the most complicated SVE phase, with the spinors being superpositions of all four basis spinors. Fig.~\ref{Fig: 1}(f) shows the order parameters along the black horizontal line marked on Fig.~\ref{Fig: 1}(b). Now one observes a second-order transition from the VPAF phase into the SVE2 phase, followed by a first-order transition into the BO phase. Then one sees a second-order transition into the BOCAF phase, and then a re-entrant first-order transition into SVE2, followed by second-order transitions into SVE4 and finally BOFM. Fig.~\ref{Fig: 1}(g) shows the evolution of the order parameters along the black horizontal line shown in Fig.~\ref{Fig: 1}(c). The system starts in SVE5, enters SVE2 via a first-order phase transition, then goes into SVE4 via a second-order phase transition, and then finally into the BO phase via a second-order transition. Fig.~\ref{Fig: 1}(h) presents the order parameter evolution on the black horizontal line of Fig.~\ref{Fig: 1}(d). The system starts in the V/SAF phase, goes through the CAF phase, enters the VPFM phase, and finally the BOFM phase. Only the final transition is first-order.

The natural expectation is that a phase transition from a phase with all order parameters nonzero to one in which some vanish should be second-order. This expectation is indeed borne out in most of the phase transitions we observe. The one phase we do not fully understand is SVE2, which has no simple description in terms of spinors. Both the integer and fractional spinors are linear combinations of all four basis spinors, in any simple basis we can think of. Furthermore, despite having all order parameters nonzero, this phase has first-order transitions to most other adjacent phases, except for SVE4. 
Now we turn to the phase diagrams for $(1,\frac{2}{5},0,0)$ where we present the type II section, keeping $g_{z\perp},~g_{\perp\perp}$ fixed while presenting the phase diagrams as functions of $g_{\perp z},~g_{zz}$.
\begin{widetext}

\begin{figure}[H]
    \centering
  \includegraphics[width=1.05\textwidth,height=0.4\textwidth]{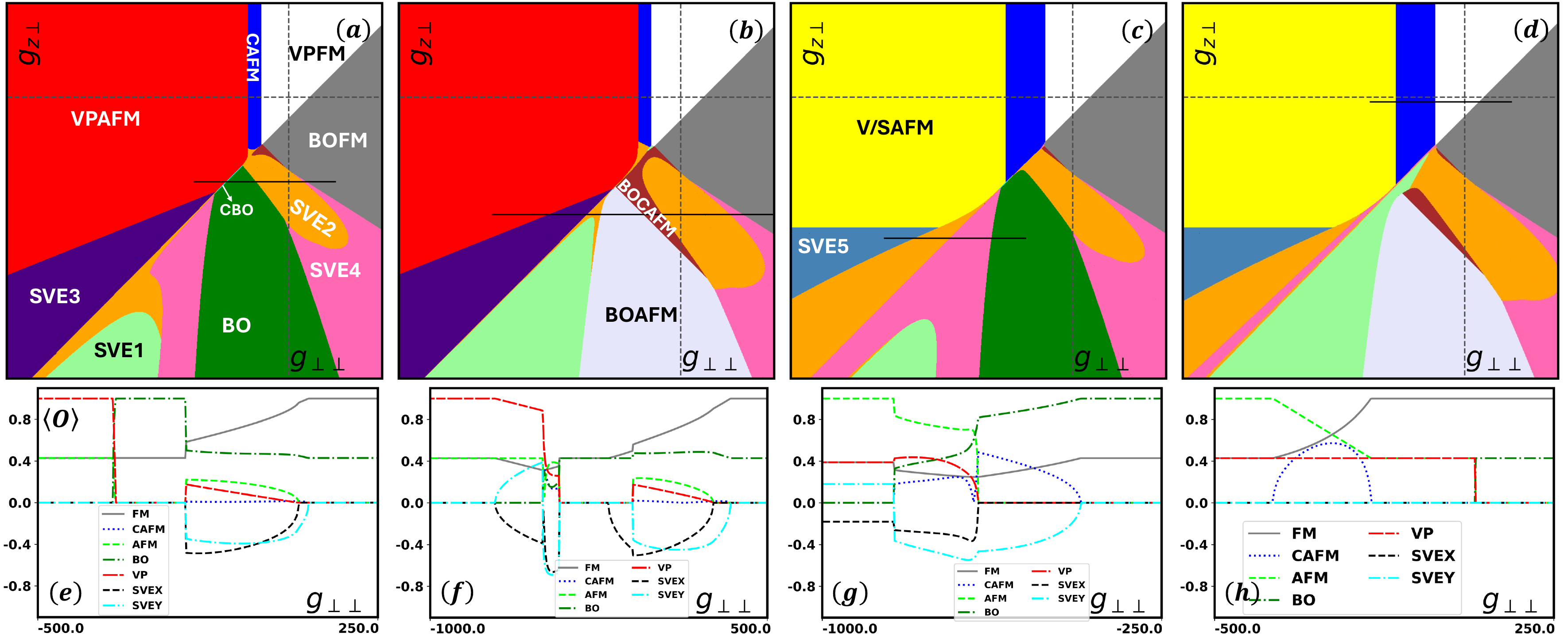}
    \caption{  Phase diagrams for $(1,\frac{2}{5},0,0)$ with $ g_{\perp\perp}\in [-1500,500]\text{meV}\cdot \text{nm}^{2}$ and $ g_{z\perp}\in [-1500,500]\text{meV}\cdot \text{nm}^{2}$. This is a type I section ($(g_{\perp z},g_{zz})$  are fixed at the values in Eq.\eqref{g_values}) with $(g_{\perp 0},g_{z0})$ being very small ($\sim 10\text{meV}\cdot \text{nm}^{2}$) and fixed (a) $g_{\perp 0}<0,\ g_{z0}<0$, (b) $g_{\perp 0}<0,\ g_{z0}>0$, (c) $g_{\perp 0}>0,\ g_{z0}<0$, (d) $g_{\perp 0}>0,\ g_{z0}>0$.  (e)-(h) Order parameters the solid black lines in (a)-(d). From left to right, the phases occurring along the horizontal sections are (e) VPAFM $\Rightarrow$ CBO $\Rightarrow$ BO $\rightarrow$ SVE2 $\Rightarrow$ SVE4 $\Rightarrow$ BOFM, (f) VPAFM $\Rightarrow$ SVE3 $\rightarrow$ SVE2 $\rightarrow$ BOAFM $\Rightarrow$ BOCAFM $\rightarrow$ SVE2 $\Rightarrow$ SVE4 $\Rightarrow$ BFM, (g) SVE5 $\rightarrow$ SVE2 $\Rightarrow$ SVE4 $\Rightarrow$ BO, (h) V/SAF $\Rightarrow$ CAFM $\Rightarrow$ VPFM $\rightarrow$ BOFM. where '$\rightarrow$' denotes a $1^{st}$-order phase transition while '$\Rightarrow$' denotes a $2^{nd}$-order phase transition.} 
    \label{Fig: 1}
\end{figure}

\begin{figure}[H]
    \centering
  \includegraphics[width=1.05\textwidth,height=0.25\textwidth]{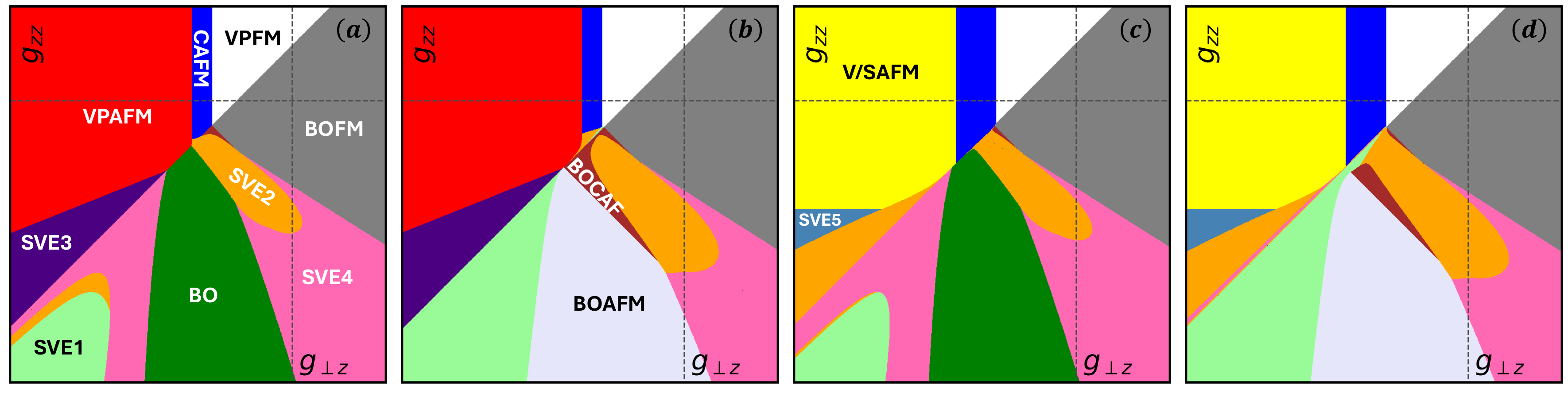}
    \caption{Type II section phase diagrams for $(1,\frac{2}{5},0,0)$   with $ g_{\perp z}\in [-3000,1000]\text{meV}\cdot \text{nm}^{2}$ and $ g_{zz}\in [-3000,1000]\text{meV}\cdot \text{nm}^{2}$. $(g_{\perp\perp},g_{z\perp})$  are fixed at the values in Eq.\eqref{g_values}, while $(g_{\perp 0},g_{z0})$ are very small ($\sim 10\text{meV}\cdot \text{nm}^{2}$) and fixed. (a) $g_{\perp 0}<0,\ g_{z0}<0$, (b) $g_{\perp 0}<0,\ g_{z0}>0$, (c) $g_{\perp 0}>0,\ g_{z0}<0$, (d) $g_{\perp 0}>0,\ g_{z0}>0$.}
    \label{Fig: 2}
\end{figure}
\end{widetext}

As expected from the discussion after Eq.~\eqref{eq: Hspeudo_n=1}, the same phases as in the type I section appear in roughly the same places in the type II section, with a few quantitative changes to the phase boundaries. We have not presented the evolution of the order parameters for these phase diagrams because they do not have any additional information than the ones in Fig.~\ref{Fig: 1}.

Now we turn to the phase diagrams for the filling $(1,\frac{2}{3},0,0)$. We will follow our procedure of showing two types of sections. For each phase diagram, we fix the nominally vanishing couplings $g_{z0},~g_{\perp0}$ at small values, and allow each to be positive or negative, which gives a set of four phase diagrams. 

In Fig.~\ref{Fig: 3} we present the type I section ($g_{zz},~g_{\perp z}$ fixed at the microscopic estimate~\cite{Wei_Xu_Sodemann_Huang_LLM_SU4_breaking_MLG_2024}). The one noteworthy difference between $\nu_1{=}(1,\frac{2}{3},0,0)$ and $\nu_2{=}(1,\frac{2}{5},0,0)$ is that the phase SVE6 appears in $\nu_1$ but not $\nu_2$. Instead, in $\nu_2$ we have SVE5, which is a special case of SVE6. It also appears that the BOFM phase does not appear at all for $(1,\frac{2}{3},0,0)$, but this is an artifact of the numerics. For this fraction, the VPFM and BOFM phases are exactly degenerate in the absence of $E_v$. To make the numerical results unique, we impose a tiny $E_V{\simeq}10^{{-}6}$ which favors the VPFM phase. In actual fact, the entire VPFM phase is degenerate with the BOFM phase for $(1,\frac{2}{3},0,0)$. Otherwise, there are quantitative changes in the phase boundaries: SVE2 shrinks considerably, while the CAF and BOCAFM phases expand. Last, but not least, SVE1 occurs mostly in the fourth quadrant for $(1,2/3,0,0)$, whereas for $(1,2/5,0,0)$ it is present only in the third quadrant. Panels (e)-(h) of Fig.~\ref{Fig: 3} show the evolution of the order parameters along selected sections of the phases diagrams in panels (a)-(d). 

Next, we present the  type II section for $(1,\frac{2}{3},0,0)$ in Fig.~\ref{Fig: 4}. Once again, as expected from the discussion after Eq.~\eqref{eq: Hspeudo_n=1}, there is not much change in the phase diagram. 
\begin{widetext}

\begin{figure}[H]
    \centering
  \includegraphics[width=1.05\textwidth,height=0.4\textwidth]{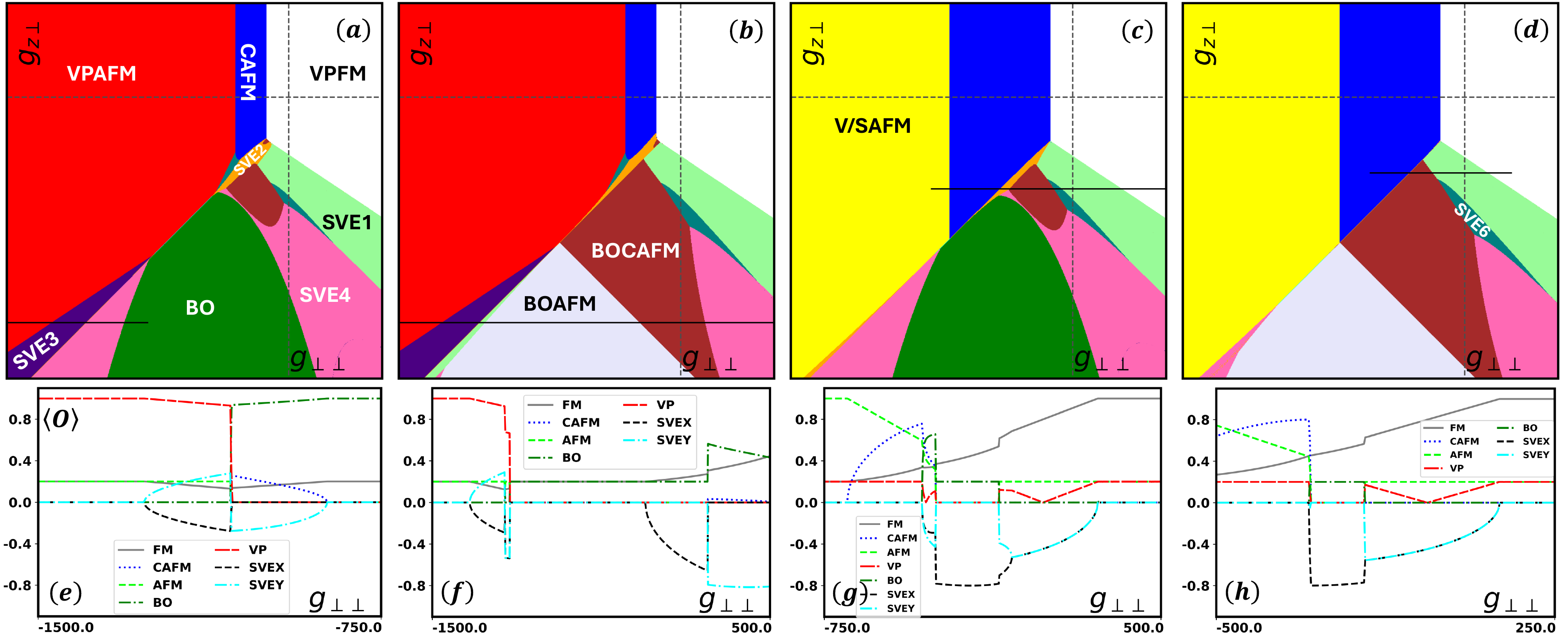}
    \caption{Type I section for phase diagrams for $(1,\frac{2}{3},0,0)$   with $ g_{\perp\perp}\in [-1500,500]\text{meV}\cdot \text{nm}^{2}$ and $ g_{z\perp}\in [-1500,500]\text{meV}\cdot \text{nm}^{2}$. $(g_{\perp z},g_{zz})$  are fixed at the values in Eq.\eqref{g_values}, with $(g_{\perp 0},g_{z0})$ being very small ($\sim 10\text{meV}\cdot \text{nm}^{2}$) and fixed. (a) $g_{\perp 0}<0,\ g_{z0}<0$, (b) $g_{\perp 0}<0,\ g_{z0}>0$, (c) $g_{\perp 0}>0,\ g_{z0}<0$, (d) $g_{\perp 0}>0,\ g_{z0}>0$.  (e)-(h) Order parameters along the solid black lines in (a)-(d). From left to right, the phases occuring on the horizontal sections are (e) VPAFM $\Rightarrow$ SVE3 $\rightarrow$ SVE4 $\Rightarrow$ BO, (f) VPAFM $\Rightarrow$ SVE3 $\rightarrow$ SVE1 $\rightarrow$ BOAFM $\Rightarrow$ BOCAFM $\rightarrow$ SVE4, (g) V/SAF $\Rightarrow$ CAFM $\rightarrow$ SVE2 $\rightarrow$ BOCAFM $\rightarrow$ SVE6 $\Rightarrow$ SVE1 $\Rightarrow$ VPFM, (h) CAF $\rightarrow$ BOCAFM $\rightarrow$ SVE1 $\Rightarrow$ VPFM, where '$\rightarrow$' denotes a $1^{st}$-order phase transition while '$\Rightarrow$' denotes a $2^{nd}$-order phase transition. }
    \label{Fig: 3}
\end{figure}

\begin{figure}[H]
    \centering
  \includegraphics[width=1.05\textwidth,height=0.25\textwidth]{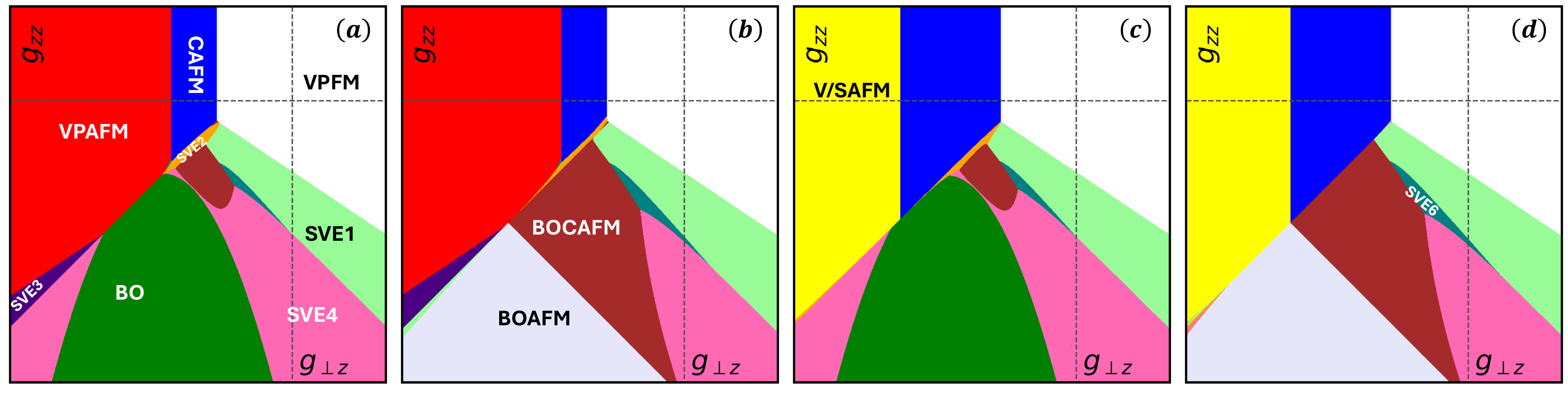}
    \caption{$(1,\frac{2}{3},0,0)$  phase diagrams with $ g_{\perp z}\in [-2000,2000]\text{meV}\cdot \text{nm}^{2}$ and $ g_{zz}\in [-3000,1000]\text{meV}\cdot \text{nm}^{2}$. $(g_{\perp\perp},g_{z\perp})$  are fixed at the values in Eq.\eqref{g_values} while $(g_{\perp 0},g_{z0})$ at very small ($\sim 10\text{meV}\cdot \text{nm}^{2}$) and fixed a) $g_{\perp 0}<0,\ g_{z0}<0$, b) $g_{\perp 0}<0,\ g_{z0}>0$, c) $g_{\perp 0}>0,\ g_{z0}<0$, d) $g_{\perp 0}>0,\ g_{z0}>0$.}
    \label{Fig: 4}
\end{figure}

\begin{figure}[H]
    \centering
  \includegraphics[width=1.05\textwidth,height=0.25\textwidth]{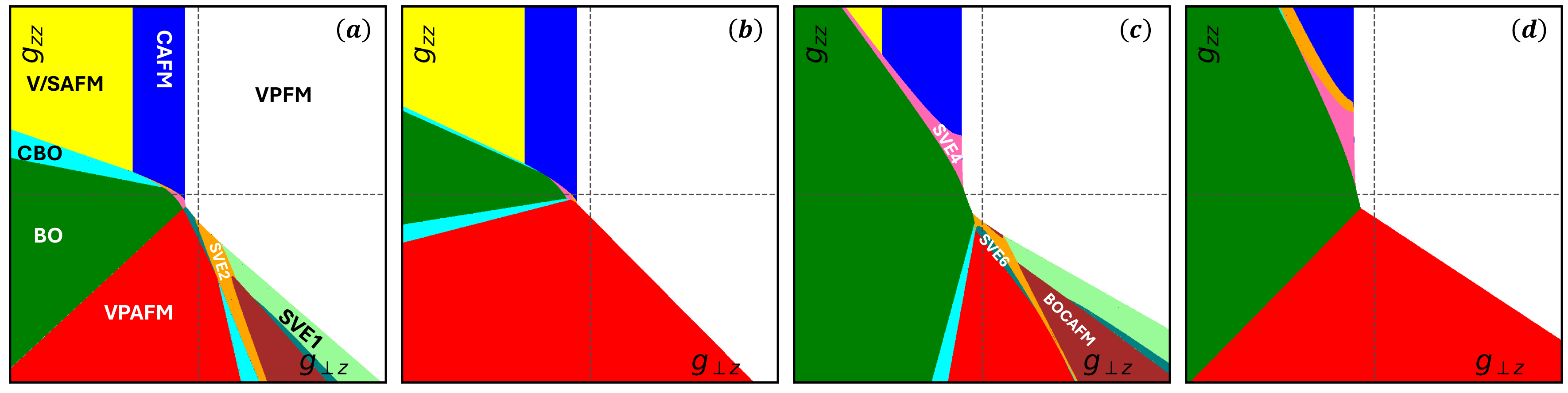}
    \caption{
    Phase diagrams for the filling $(1,\frac{2}{3},0,0)$ in the ZLLs with $ g_{\perp z}\in [-1600,1600]\text{meV}\cdot \text{nm}^{2}$ and $ g_{zz}\in [-1600,1600]\text{meV}\cdot \text{nm}^{2}$. The $m{=}1$ Haldane pseudopotentials of each $g_{ij}$ are $0.2$ times the corresponding $m=0$ pseudopotentials in magnitude. (a) $u^{(1)}_\perp=-0.2u^{(0)}_\perp,~u^{(1)}_z=-0.2u^{(0)}_z$. (b) $u^{(1)}_\perp=0.2u^{(0)}_\perp,~u^{(1)}_z=-0.2u^{(0)}_z$. (c) $u^{(1)}_\perp=-0.2u^{(0)}_\perp,~u^{(1)}_z=0.2u^{(0)}_z$. (d) $u^{(1)}_\perp=0.2u^{(0)}_\perp,~u^{(1)}_z=0.2u^{(0)}_z$. Fewer phases appear than in the $\bn=1$ LL manifold. The SVE phases shrink, and the BOAFM phase does not appear. The broad outlines of the phase diagram are also very different. For example, the VPAFM and BO phases seem to have exchanged places. }
    \label{Fig: 5}
\end{figure}
\end{widetext}

Finally, to end this section and to draw a contrast between our current results and the results in the ZLLs, we show the phase diagram for natural couplings in the ZLL for the filling $(1,\frac{2}{3},0,0)$. From Eq.~\eqref{n0_haldane}, which we reproduce below for convenience,
\begin{equation}
u^{(0)}_{\perp}= \frac{g_{\perp 0} + g_{\perp z}}{4 \pi \ell^2},\ 
u^{(0)}_{z}= \frac{g_{z 0} + g_{z z}}{4 \pi \ell^2}.
\end{equation}
We see that the couplings $g_{i\perp }$ are not relevant to the ZLLs. We will keep $g_{i0}{=}0$ (the value of the microscopic estimate~\cite{Wei_Xu_Sodemann_Huang_LLM_SU4_breaking_MLG_2024}) and vary $g_{\perp z},~g_{zz}$, as in the type II sections above.  To get the full set of phases appropriate to natural couplings in the ZLLs we have to go beyond USR interactions and introduce Haldane pseudopotentials for the $m{=}1$ channel $u^{(1)}_{i}$. For illustrative purposes, we will use $u^{(1)}_i{=}{\pm} 0.2 u^{(0)}_i$ Fig.~\ref{Fig: 5} shows the phase diagram for natural couplings for the filling $(1,\frac{2}{3},0,0)$ in the ZLL for the cases (a) $u^{(1)}_\perp{=}{-}0.2u^{(0)}_\perp,~u^{(1)}_z{=}{-}0.2u^{(0)}_z$; (b) $u^{(1)}_\perp{=}{-}0.2u^{(0)}_\perp,~u^{(1)}_z{=}0.2u^{(0)}_z$; (c) $u^{(1)}_\perp{=}0.2u^{(0)}_\perp,~u^{(1)}_z{=}{-}0.2u^{(0)}_z$; (d) $u^{(1)}_\perp{=}0.2u^{(0)}_\perp,~u^{(1)}_z{=}0.2u^{(0)}_z$. Note that in previous work by three of us~\cite{Jincheng2024}, we plotted the phase digrams differently, keeping $u^{(1)}_i$ fixed while varying  $u^{(0)}_i$. We use the current parameterization as it is similar to what occurs in the $\bn{\neq}0$ LL manifolds, where all $u^{(m)}$'s scale with the $g_{ij}$.

Far fewer phases appear as compared to Fig.~\ref{Fig: 4}, and the phases that do appear do so in different regions of the phase diagram than in the $n{=}1$ LL manifold. The SVE phases seem to have a much-reduced domain of stability. As pointed out in the introduction, the reason for the differences is the large values of the Hartree parts of each coupling in the ZLL and the near vanishing of the Hartree parts in the $\bn{\neq}0$ LL manifolds.

\subsection{Phases diagrams for three-component singlet states}
The situation for 3CS is even more complex than for the 2CS because now three spinors are involved. We divide the phases into three classes as for the 2CS: In the first class all spinors are valley polarized, in the second class they are valley equatorial, while the third class has spin-valley entangled spinors.  It is no longer possible for all three spinors to have the same spin polarization. However, we will continue to call the state with the maximum possible $S_z$ a ferromagnet, while the state with minimal $S_z$, with all spinors having definite $S_z{=}{\pm}\frac{1}{2}$, will be labeled an antiferromagnet. In all, we find twelve distinct phases in the $\bn\neq0$ LL manifolds, as opposed to the eight phases found in the ZLL manifold. 

Below, we list the different phases and the corresponding occupied spinors. The first spinor is fully occupied, while the next two are each fractionally occupied, and in a singlet state.

\begin{itemize}[left=0pt]
    \item VPFM1: $|K,\uparrow\rangle,\ |K^\prime,\uparrow\rangle,\ |K^\prime,\downarrow\rangle$
    \item VPFM2: $|K,\uparrow\rangle,\ |K,\downarrow\rangle,\ |K^\prime,\uparrow\rangle$
    \item VPAFM: $|K,\uparrow\rangle$, $|K,\downarrow\rangle$, $|K^\prime,\downarrow\rangle$
    \item VPCAFM: $|K,\bs_a\rangle,\ |K,-\bs_a\rangle,\ |K^\prime,\bs_b\rangle$
    \item BOFM1: $|-\hat e_x,\uparrow\rangle,\ |\hat e_x,\uparrow\rangle,\ |\hat e_x,\downarrow\rangle$
    \item BOFM2: $|-\hat e_x,\uparrow\rangle,\ |\hat e_x,\uparrow\rangle,\ |-\hat e_x,\downarrow\rangle$
    \item CBO: $|-\btau_a,\uparrow\rangle,\ |\btau_a,\uparrow\rangle,\ |\btau_b,\downarrow\rangle$
    \item SVE1: $|K,\uparrow\rangle,\ |K^\prime,\downarrow\rangle, \cos\frac{\alpha}{2}|K,\downarrow\rangle+\sin\frac{\alpha}{2}|K^\prime,\uparrow\rangle$
    \item SVE2: There is no simple description for this phase, all order parameters are nonzero.
    \item SVE3: $|\hat e_x, \bs_a\rangle,\ |\hat e_x,-\bs_a\rangle,\ |-\hat e_x,\bs_b\rangle\quad$
    \item SVE4: \\
    $\cos\frac{\alpha_1}{2}|\hat e_x,\downarrow\rangle+\sin\frac{\alpha_1}{2}|-\hat e_x,\uparrow\rangle$,\\
$\sin\frac{\alpha_1}{2}|\hat e_x,\downarrow\rangle-\cos\frac{\alpha_1}{2}|-\hat e_x,\uparrow\rangle$,\\
$\cos\frac{\alpha_2}{2}|\hat e_x,\uparrow\rangle+\sin\frac{\alpha_2}{2}|-\hat e_x,\downarrow\rangle$
    \item B/VAFM: $|-\hat e_x,\uparrow\rangle,\ |K,\downarrow\rangle,\ |K^\prime,\downarrow\rangle$    
\end{itemize}
 All the nontrivial angles in these spinors can be determined by minimization of the energy functional. The last phase on the list, B/VAFM appears not to belong to any class. However, it is only the projector to the singlet subspace that matters and not the specific spinors that we use to describe them. The fractional spinors form a singlet in the valley space and can be described by any pair of orthonormal vectors in the valley space. With this, the B/VAFM phase turns out to be a special case of SVE3 with ${\hat e}_x\rightarrow -{\hat e}_x,~\bs_a=\uparrow,~\bs_b=\downarrow$.

\begin{figure*}
    \centering
  \includegraphics[width=1.05\textwidth,height=0.4\textwidth]{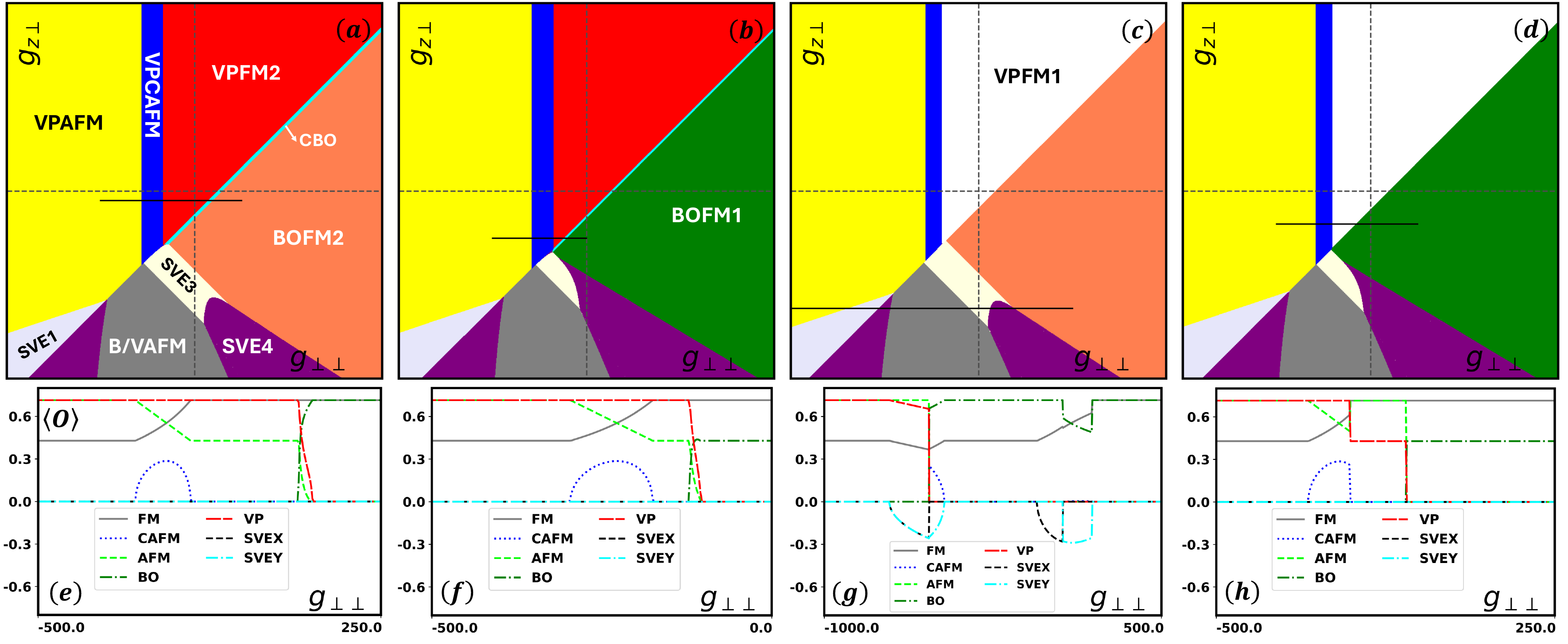}
    \caption{Type I phase diagrams for $(1,[\frac{1}{5},\frac{1}{5}],0)$  with $ g_{\perp\perp}\in [-1000,1000]\text{meV}\cdot \text{nm}^{2}$ and $ g_{z\perp}\in [-1000,1000]\text{meV}\cdot \text{nm}^{2}$. $(g_{\perp z},g_{zz})$  are fixed at the values in Eq.\eqref{g_values} while $(g_{\perp 0},g_{z0})$ are kept very small ($\sim 10\text{meV}\cdot \text{nm}^{2}$) and fixed. (a) $g_{\perp 0}<0,\ g_{z0}<0$, (b) $g_{\perp 0}<0,\ g_{z0}>0$, (c) $g_{\perp 0}>0,\ g_{z0}<0$, (d) $g_{\perp 0}>0,\ g_{z0}>0$.  (e)-(h) Order parameters $\langle O\rangle's$ along horizontal sections denoted by black solid lines in (a)-(d). From left to right, the phases occurring along the horizontal sections are (e) VPAF $\Rightarrow$ CAF $\Rightarrow$ VPFM2 $\Rightarrow$ CBO $\rightarrow$ BOFM2, (f) VPAF $\Rightarrow$ CAF $\rightarrow$ VPFM2 $\rightarrow$ CBO $\Rightarrow$ BOFM1, (g) VPAF $\rightarrow$ SVE4 $\rightarrow$ B/VAF $\Rightarrow$ SVE3 $\rightarrow$ SVE4 $\rightarrow$ BOFM2, (h) VPAF $\Rightarrow$ CAF $\rightarrow$ VPFM1 $\rightarrow$ BOFM1. As usual, '$\rightarrow$' denotes a $1^{st}$-order phase transition while '$\Rightarrow$' denotes a $2^{nd}$-order phase transition.}
    \label{Fig: 6}
\end{figure*}

In Fig.~\ref{Fig: 6}  we present the phases diagrams for the type I section at filling $\vec \nu {=}(1,[\frac{1}{5},\frac{1}{5}],0)$ ($(g_{\perp z},g_{zz})$ fixed). For the 3CS, it is evident that $g_{z0}{>}0$ stabilizes the BOFM1 phase while $g_{z0}{<}0$ stabilizes the BOFM2 phase. Similarly, $g_{\perp0}{>}0$ stabilizes the VPFM1 phase, while $g_{\perp0}<0$ stabilizes the VPFM2 phase. The SVE2 phase never appears in this type of section, presumably because the values of $(g_{\perp z},g_{zz})$ are fixed at the microscopic estimate~\cite{Wei_Xu_Sodemann_Huang_LLM_SU4_breaking_MLG_2024}. The B/VAFM, with its mix of valley polar spinors in the fractional subspace and valley equatorial spinors in the integer subspace, occupies a significant part of the phase diagram.

In the lower panels of Fig.~\ref{Fig: 6} we show the evolution of the order parameters along some representative sections of the corresponding phase diagram. Most of the phase transitions are second-order. By examining the spinors listed above it is easy to see why the transitions SVE4$\rightarrow$ B/VAFM, CAF $\rightarrow$ VPFM1, and VPFM1 $\rightarrow$ BOFM1 are first-order. 

\begin{figure*}
    \centering
  \includegraphics[width=1.05\textwidth,height=0.25\textwidth]{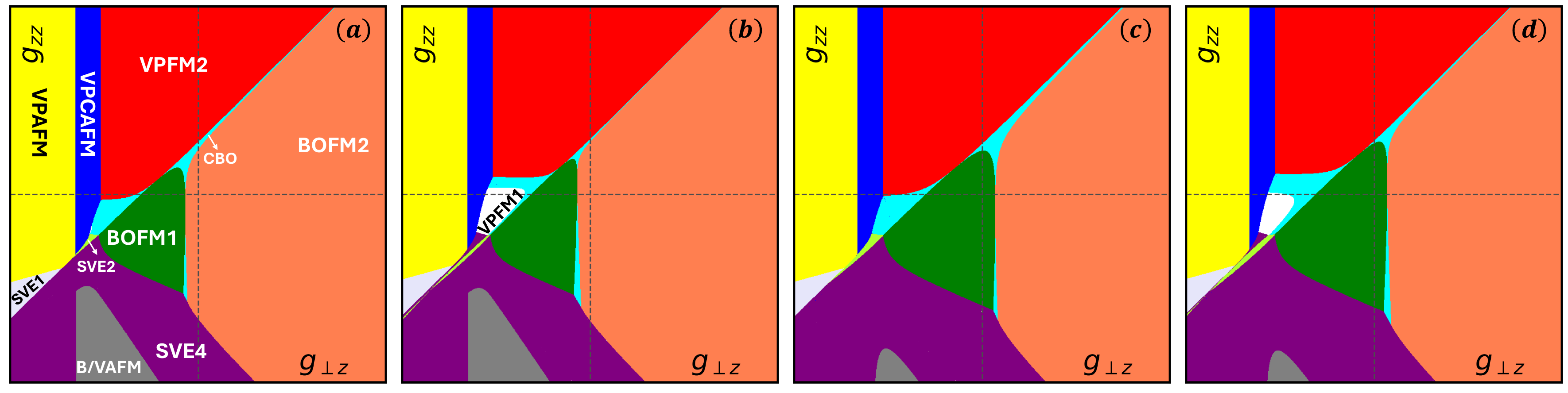}
    \caption{Type II section phase diagrams for $(1,[\frac{1}{5},\frac{1}{5}],0)$ with $ g_{\perp z}\in [-2000,2000]\text{meV}\cdot \text{nm}^{2}$ and $ g_{zz}\in [-2000,2000]\text{meV}\cdot \text{nm}^{2}$. $(g_{\perp\perp},g_{z\perp})$  are fixed at the values in Eq.\eqref{g_values} while $(g_{\perp 0},g_{z0})$ are taken to be very small ($\sim 10\text{meV}\cdot \text{nm}^{2}$) and fixed. a) $g_{\perp 0}<0,\ g_{z0}<0$, b) $g_{\perp 0}<0,\ g_{z0}>0$, c) $g_{\perp 0}>0,\ g_{z0}<0$, d) $g_{\perp 0}>0,\ g_{z0}>0$.}
    \label{Fig: 7}
\end{figure*}

In Fig.~\ref{Fig: 7} we present the type II section for $(1,[\frac{1}{5},\frac{1}{5}],0)$ ($g_{\perp\perp},~g_{z\perp}$ fixed). In contrast to the previous section for 2CS, now the phase diagrams are quite different for the type I and type II sections. This is because the $m{=}2$ Haldane pseudopotential does affect the singlet $2/5$ states. Now the SVE2 phase appears in small regions of the coupling constants. As in the type I section, one needs $g_{z0}{>}0$ to stabilize the VPFM1 phase, even though its regime of stability is now greatly reduced. The rest of the phases seem quite insensitive to the signs of $g_{z0},~g_{\perp0}$. Both BOFM1 and BOFM2 phases appear for all signs of $g_{z0},~g_{\perp0}$. 

\begin{figure*}
    \centering
  \includegraphics[width=1.05\textwidth,height=0.4\textwidth]{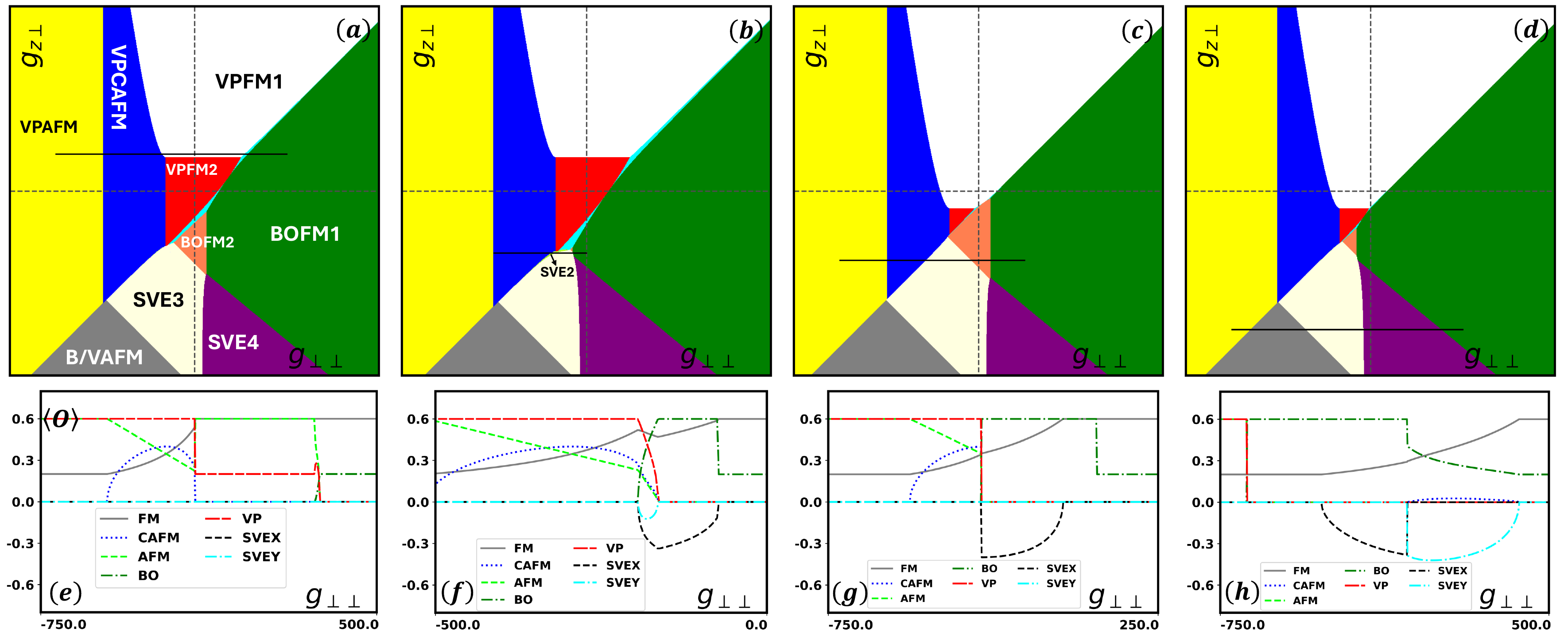}
    \caption{Type I phase diagrams for $(1,[\frac{1}{3},\frac{1}{3}],0)$   with $ g_{\perp\perp}\in [-1000,1000]\text{meV}\cdot \text{nm}^{2}$ and $ g_{z\perp}\in [-1000,1000]\text{meV}\cdot \text{nm}^{2}$. $(g_{\perp z},g_{zz})$  are fixed at the values in Eq.\eqref{g_values} while $(g_{\perp 0},g_{z0})$ are kept very small ($\sim 10\text{meV}\cdot \text{nm}^{2}$) and fixed. (a) $g_{\perp 0}<0,\ g_{z0}<0$, (b) $g_{\perp 0}<0,\ g_{z0}>0$, (c) $g_{\perp 0}>0,\ g_{z0}<0$, (d) $g_{\perp 0}>0,\ g_{z0}>0$.  (e)-(h) Order parameters $\langle O\rangle's$ along horizontal sections denoted by black solid lines in (a)-(d). From left to right, the phases occurring along the horizontal sections are (e) VPAF $\Rightarrow$ CAF $\rightarrow$ VPFM1 $\Rightarrow$ CBO $\Rightarrow$ BOFM1, (f) CAF $\Rightarrow$ SVE2 $\Rightarrow$ SVE3 $\rightarrow$ BOFM1, (g) VPAF $\rightarrow$ CAF $\rightarrow$ SVE3 $\Rightarrow$ BOFM2 $\rightarrow$ BOFM1, (h) VPAF $\rightarrow$ B/VAF $\Rightarrow$ SVE3 $\rightarrow$ SVE4 $\Rightarrow$ BOFM1.}
    \label{Fig: 8}
\end{figure*}

\begin{figure*}
    \centering
  \includegraphics[width=1.05\textwidth,height=0.25\textwidth]{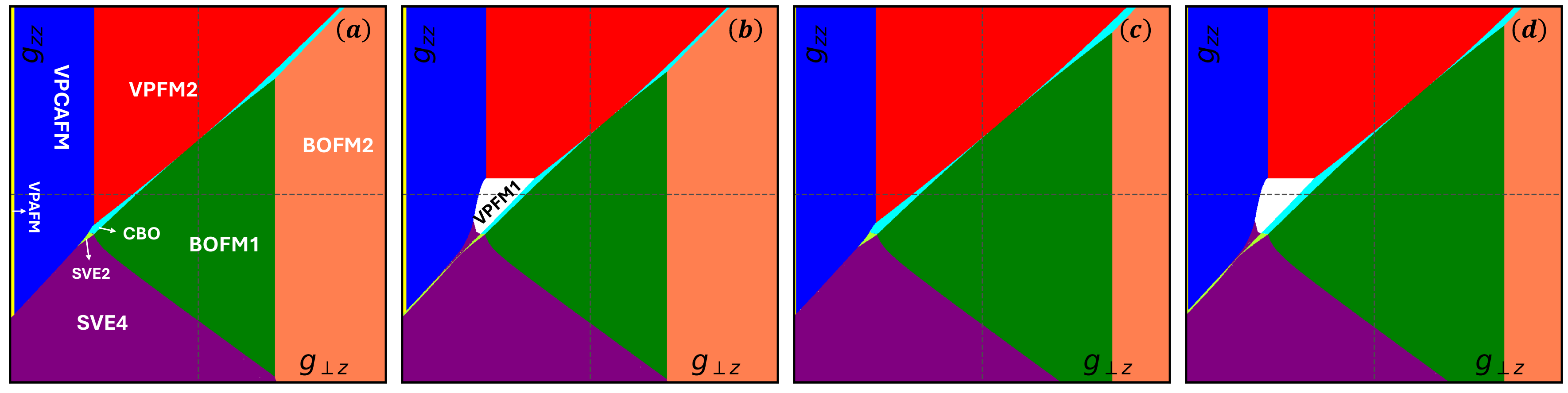}
    \caption{Type II phase diagrams for $(1,[\frac{1}{3},\frac{1}{3}],0)$  phase diagrams with $ g_{\perp z}\in [-2000,2000]\text{meV}\cdot \text{nm}^{2}$ and $ g_{zz}\in [-2000,2000]\text{meV}\cdot \text{nm}^{2}$. $(g_{\perp\perp},g_{z\perp})$  are fixed at the values in Eq.\eqref{g_values} while $(g_{\perp 0},g_{z0})$ are kept very small ($\sim 10\text{meV}\cdot \text{nm}^{2}$) and fixed. a) $g_{\perp 0}<0,\ g_{z0}<0$, b) $g_{\perp 0}<0,\ g_{z0}>0$, c) $g_{\perp 0}>0,\ g_{z0}<0$, d) $g_{\perp 0}>0,\ g_{z0}>0$.}
    \label{Fig: 9}
\end{figure*}

In Fig.~\ref{Fig: 8} we present phase diagrams for $(1,[\frac{1}{3},\frac{1}{3}],0)$ for the type I section, which is quite different from the type I sections for $(1,[\frac{1}{5},\frac{1}{5}],0)$. The reason is that the contribution to the variational energy $E_{FF}$ arising from the fractional-fractional interaction in $(1,[\frac{1}{3},\frac{1}{3}],0)$ is sensitive to both the $m{=}1$ and $m{=}2$ Haldane pseudopotentials, whereas in $(1,[\frac{1}{5},\frac{1}{5}],0)$ $E_{FF}$ is only sensitive to the $m{=}2$ pseudopotentials. Recall that in the type I section, the $m{=}2$ pseudopotentials remain constant while the $m{=}0,1$ vary. In Fig.~\ref{Fig: 9}, we present phase diagrams for $(1,[\frac{1}{3},\frac{1}{3}],0)$ for the type II section where m=2 pseudopotentials are not fixed. As in the case of the filling $(1,[\frac{1}{5},\frac{1}{5}],0)$, the type I and type II sections are quite different, because these are 3CS, for which the intra-fractional contribution to the variational energy, $E_{FF}$, does depend on the $m=1,2$ pseudopotentials.

\begin{figure*}
    \centering
  \includegraphics[width=1.05\textwidth,height=0.25\textwidth]{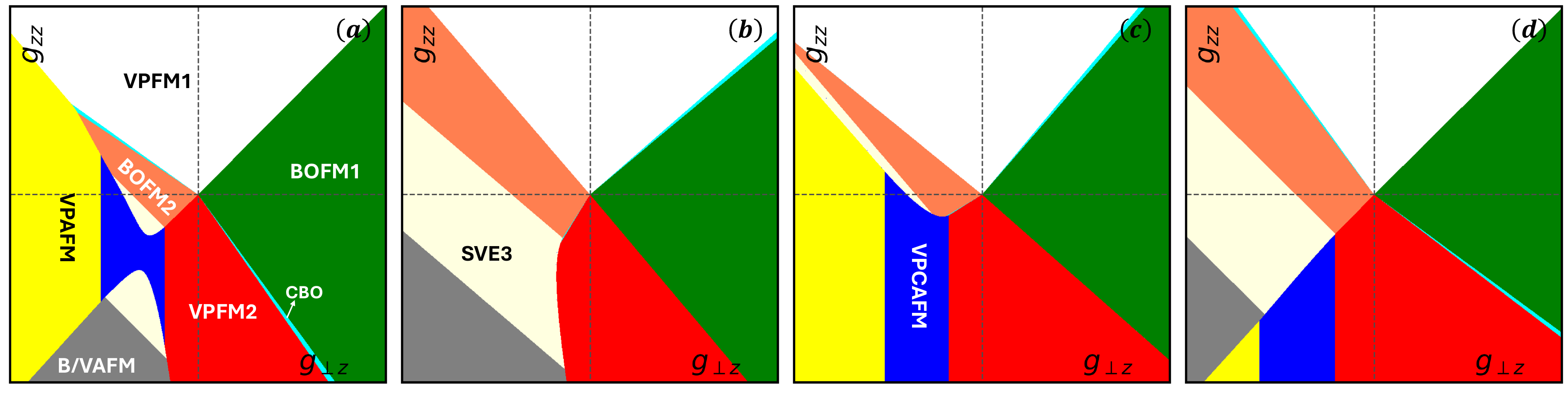}
    \caption{Type II section phase diagrams for $(1,[\frac{1}{3},\frac{1}{3}],0)$  in the ZLLs,  with $ g_{\perp z}\in [-800,800]\text{meV}\cdot \text{nm}^{2}$ and $ g_{zz}\in [-800,800]\text{meV}\cdot \text{nm}^{2}$. The $m{=}1$ Haldane pseudopotentials of each $g_{ij}$ are $0.1$ times the corresponding $m=0$ pseudopotentials in magnitude. (a) $u^{(1)}_\perp=-0.1u^{(0)}_\perp,~u^{(1)}_z=-0.1u^{(0)}_z$. (b) $u^{(1)}_\perp=0.1u^{(0)}_\perp,~u^{(1)}_z=-0.1u^{(0)}_z$. (c) $u^{(1)}_\perp=-0.1u^{(0)}_\perp,~u^{(1)}_z=0.1u^{(0)}_z$. (d) $u^{(1)}_\perp=0.1u^{(0)}_\perp,~u^{(1)}_z=0.1u^{(0)}_z$. Except for the SVE2 and SVE4 phases, all the phases appearing in the type II section in the $\bn=1$ manifold for $(1,[\frac{1}{3},\frac{1}{3}],0)$ appear here as well, but the topology of the phase diagram is quite different. }
    \label{Fig: 10}
\end{figure*}

To end this subsection, we compare a representative set of phase diagrams for $(1,[\frac{1}{3},\frac{1}{3}],0)$ in the ZLLs (Fig.~\ref{Fig: 10}) to the type II section found above in the $\bn{=}1$ LL manifold (Fig.~\ref{Fig: 9}). The SVE2 and SVE4 phases are missing. We recall that the SVE4 phase never appears in the ZLLs~\cite{Jincheng2024} but the SVE2 phase does. We do not see SVE2 in Fig.~\ref{Fig: 10} because of the small magnitude of $u_i^{(1)}$ compared to $u_i^{(0)}$ chosen here. The B/VAFM phase is a special case of SVE3, which does appear in our previous work~\cite{Jincheng2024}.

\section{Conclusions, caveats, and open questions}
\label{sec: discussion_summary}
Graphene has proven to be a fertile source of quantum Hall ferromagnets, both at integer and fractional fillings. Our focus in this work has been on fractional fillings near half-filling of Landau level manifolds other than $\bn{=}0$, which was the topic of a previous work by three of us~\cite{Jincheng2024}. Since the long-range Coulomb interaction is fully symmetric in the spin/valley labels, the residual short-range anisotropic interactions choose the ground state. We find a huge diversity of phases at these fractions, with different phases being stabilized in various regions of the space of short-range interactions. 

The symmetry-allowed anisotropic short-range couplings have been classified~\cite{RG1_Graphene_Aleiner_2007, RG2_Graphene_Aleiner_2008}, and their strengths estimated recently~\cite{Wei_Xu_Sodemann_Huang_LLM_SU4_breaking_MLG_2024}. There are six independent couplings, two of which ($g_{z0},~g_{\perp 0}$) happen to vanish at the leading order in the microscopic estimate~\cite{Wei_Xu_Sodemann_Huang_LLM_SU4_breaking_MLG_2024}. 

We use the fact that, despite the strongly correlated nature of the states, the variational energy can be expressed in a form very similar to the Hartree-Fock (HF) approximation~\cite{Jincheng2024} used for integer states. The only subtlety is that certain coefficients appearing in front of the HF terms for each Haldane pseudopotential $V_m$ depend on $m$ and on the nature of the fractional state itself. These coefficients have to be extracted numerically.

There are several differences between the $\bn{=}0$ Landau level manifold (the ZLLs) and the higher LL manifolds. Firstly, even if one assumes that the short-range anisotropies are ultra-short-range on the scale of the magnetic length (as we do in most of this work), the projection to any LL manifold with $\bn{\neq}0$ will generate effective longer-range couplings~\cite{Stefanidis_Sodemann2022, Stefanidis_Sodemann2023} due to the form factors induced by the singlet-particle wave functions [Eq.~\eqref{eq: um_B}]. These are represented as Haldane pseudopotentials $V_m$ for $m{>}0$. Secondly, the Hartree parts of the couplings in the $\bn{\neq}0$ LL manifolds vanish in the microscopic estimate~\cite{Wei_Xu_Sodemann_Huang_LLM_SU4_breaking_MLG_2024} [Eq.~\eqref{eq: H and F XXZ couplings}], whereas they are the largest contributions in the ZLLs. This means the ``natural" values of the anisotropic couplings are quite different between the ZLLs and the higher LL manifolds. Finally, wave functions in each valley in the $\bn{\neq}0$ LL manifolds have equal weight on each sublattice [Eq.~\eqref{eq: nLL wave functions}]. Consequently, sublattice symmetry breaking due to alignment with the encapsulating hBN~\cite{KV_HBN_Jung2015, KV_HBN_Jung2017} is ineffective in producing a valley Zeeman coupling in the $\bn{\neq}0$ manifolds, in contrast to the situation in the $\bn{=}0$ manifold, where sublattice (equivalent to valley in the ZLLs) symmetry breaking can be the largest scale~\cite{KV_DGG_2013, STM_Yazdani2021visualizing}. 

The fractional states we consider are either two-component states (2CSs), with one integer-filled spinor and one fractionally filled spinor, or three-component states (3CSs), with one integer-filled spinor and two fractionally filled spinors forming a singlet FQH state. The phases stabilized by the short-range interactions are quite distinct in these two cases. 

For the 2CS in the $\bn{=}1$ LL manifold, we find a total of 15 different phases, a considerably larger variety than the 10 phases found by a subset of us in earlier work in the ZLLs~\cite{Jincheng2024}. The difference between the phase diagrams in the ZLLs and the $\bn{\neq}0$ LL manifolds results from the fact that the Hartree parts $u_{iH}$ of the couplings [Eq.~\eqref{eq: H and F XXZ couplings}] depend only $g_{z0},g_{\perp0}$. These nominally vanish in the microscopic estimate~\cite{Wei_Xu_Sodemann_Huang_LLM_SU4_breaking_MLG_2024}, leading to small $u_{iH}$ for $\bn\neq0$, whereas $u_{iH}$ are large for $\bn{=}0$. We give small values to $g_{i0}$ to break the near-degeneracies between valley-polarized states and intervalley coherent states. One can think of the states we find as being broadly divided into three classes. In the first class, the occupied spinors can be written as direct products $|\btau\rangle{\otimes}|\bs\rangle$, where $\btau{=}{\pm}{\hat e}_z{=}{\bf K}/{\bf K}'$. The spinors are thus valley polarized (VP) and we prefix the corresponding phases by VP. In the second class, the spinors can also be written as direct products, but now $\btau$ lies in the equatorial plane, signifying intervalley coherence, which we represent as bond-ordered (BO).  The third class has occupied spinors that cannot be written as direct products. We call such phases spin-valley entangled (SVE). Note that in the ZLLs we had a correspondence between VP and CDW states due to valley-sublattice locking, but this is no longer the case for $\bn{\neq} 0$ LL manifolds. For the same reason, when the two valleys have opposite spin polarization, in the ZLLs we can identify this with lattice antiferromagnetism, but for $\bn{\neq}0$ this is no longer true. 

Coming to specifics for the 2CS, in the $\bn{=}1$ LL manifold we see two ferromagnetic phases (VPFM and BOFM), two types of antiferromagnets (VPAFM and BOAFM), and two types of canted antiferromagnets (CAFM and BOCAFM). Specific signs of $g_{z0},~g_{\perp0}$ (nominally vanishing in the microscopic estimate~\cite{Wei_Xu_Sodemann_Huang_LLM_SU4_breaking_MLG_2024}) stabilize the valley polar and bond-ordered states. The phase diagrams for $(1,\frac{2}{5},0,0)$ and $(1,\frac{2}{3},0,0)$ are rather similar, with a few quantitative differences in the locations of the phase boundaries and domains occupied by the phases. However, there is a striking contrast between these fillings in the $\bn{=}1$ LL manifold and our earlier $\bn{=}0$ (ZLL) results. The primary reason is the near vanishing of the Hartree parts of the two couplings in $\bn{\neq}0$ LL manifolds. This difference between the ZLLs and higher LL manifolds is one of the main messages of this work.

Let us now turn to our results for 3CS states. We find twelve different phases in the $\bn{=}1$ LL manifold, as compared to the eight phases~\cite{Sodemann_MacDonald_2014, Jincheng2024} found in previous work in the ZLLs. As in the case of the 2CS, one can classify the states broadly as having occupied spinors that are either valley-polar, valley-equatorial or spin-valley entangled. We find four different ferromagnetic states (defined as having the maximum possible spin polarization), an antiferromagnetic state, a canted antiferromagnetic state, a bond-ordered state, and five spin-valley entangled states.  Once again, the phase diagrams are quite rich, with the signs of the couplings $g_{i0}$ helping to choose between different types of ferromagnets. The phase diagrams of $(1,[\frac{1}{5},\frac{1}{5}],0)$ are quite different from the corresponding phase diagrams of $(1,[\frac{1}{3},\frac{1}{3}],0)$, mainly because the $m{=}1$ Haldane pseudopotentials affect the latter more strongly. As in the 2CS, the ZLL phase diagrams for $(1,[\frac{1}{3},\frac{1}{3}],0)$ are quite different from those in the $\bn{=}1$ LL manifold. 

Overall, there are two messages we would like to emphasize: (i) The phase diagrams of FQH states in the ZLL are quite different from those in $\bn{\neq}0$ LL manifolds for natural couplings. There is no qualitative difference in the phase diagrams in any $\bn{\neq}0$ manifold, though there will be quantitative changes. (ii) All the SVE phases spontaneously break both magnetic and lattice $U(1)$ symmetries simultaneously. The consequence of breaking the spin $U(1)$ is the presence of a Goldstone mode in the bulk, which could be detected either by antiferromagnon transport~\cite{JunZhu_2021} or by noise measurements~\cite{Anindya_2024}. However, because the valley $U(1)$ symmetry is broken down to a $Z_3$ when higher-fermion interactions are included, the spontaneous breaking of the valley $U(1)$ is not expected to lead to a Goldstone mode. 
There are several open questions. One of the most important concerns is how to identify a given phase. Unlike in the ZLL, where valley polarization is manifested as a CDW order due to valley-sublattice locking, in the $\bn{\neq}0$ LL manifolds, the situation is more subtle. Both valleys have an equal superposition of both sublattices, so valley polarization does not lead to CDW order. Similarly, in the ZLL, spontaneous breaking of the $U(1)_v$ symmetry leads to bond order. In the $\bn{\neq}0$ LL manifolds it is still true that breaking the $U(1)_v$ symmetry leads to breaking the lattice symmetry since a new reciprocal lattice vector has been introduced. However, the precise way in which this lattice symmetry breaking manifests itself is an open question. Coming to the spin-valley entangled phases, we are not aware of any probe that could directly detect this kind of order, so the design of such a probe is a pressing issue. 

Let us briefly recapitulate the assumptions we have made in our analysis, and how one might try to go beyond them. We have completely ignored the effects of disorder and nonzero temperature. Disorder will tend to favor states that break translation invariance over those that do not. Thus, we expect that all the bond-ordered states will expand at the expense of their neighbors if disorder is present. An important effect of nonzero temperature is that any phase that spontaneously breaks a $U(1)$ symmetry at $T{=}0$ will only have power-law order at $T{>}0$  as a consequence of the Hohenberg-Mermin-Wagner theorem~\cite{Mermin_Wagner_1966, Hohenberg_1967}. However, gapless modes will continue to exist until the Kosterlitz-Thouless transition is reached~\cite{KT}.

We have completely ignored the sublattice symmetry breaking $E_{SL}$ induced by the hBN substrate in the $\bn{\neq}0$ manifolds~\cite{KV_DGG_2013, KV_expt_2013, KV_HBN_Jung2015, KV_HBN_Jung2017} This is valid at leading order because the wave functions in each valley are an equal superposition on the two sublattices for $E_{SL}{=}0$. However, a more careful treatment would include the change in the one-body wave functions due to the sublattice field. These effects would appear at order $E_{SL}/(\hbar v_F/\ell)$. In real samples $E_{SL}$ can be a few $meV$~\cite{KV_DGG_2013} (and may be enhanced further by exchange effects), implying that these corrections could have a significant effect. In this context, there is an outstanding puzzle posed by the response of FQH gaps to a parallel field in the $\bn{=}1$ LL manifold~\cite{GoldhaberGordon2015}. The experiment suggests that the ground states may not be spin-polarized. However, previous work taking into account only the Coulomb interaction~\cite{Balram15c}, and the calculations in this work, show that the ground states are spin-polarized in the physical regime of couplings. The possibility that sublattice symmetry breaking could play a role in resolving this puzzle is something we intend to investigate in future work. 

Theoretically, one of the most important open questions is the effect of Landau level mixing on the effective interactions projected to a particular LL manifold. Several perturbative calculations~\cite{RG_Peterson_Nayak_2013, RG_Peterson_Nayak_2014, RG_Sodemann_MacD2013, Chunli_FRG2023} for the effective interactions in the ZLL have been carried out over the past decade. While these are likely to be quite accurate for strongly screened Coulomb interactions, they may not be adequate for weakly screened samples. A renormalization group calculation keeping all the couplings would help to clarify the issue. We look forward to addressing these and other outstanding issues in future work. 

As we have emphasized before, the $n{=}0,1$ LLs of graphene are not the only places where we can see these phases. Excitingly, recently, the zero magnetic field analog of the FQHE, namely, the fractional quantum anomalous Hall effect has been seen in twisted bilayers of transition metal dichalcogenides such as MoTe$_2$~\cite{FQAH_MoTe2_Xu_2023a, FQAH_MoTe2_Xu_2023b, FQAH_MoTe2_Mak_Shan_2023, FQAHE_MoTe2_Li_2023} and also in pentalayer graphene~\cite{FQAH_Pentalayer_Graphene_Ju_2024}. In these systems too, the Coulomb and short-range interactions should compete and can potentially realize the phases we discussed. In the future, it is worth looking into these as well as other systems such as bilayer graphene to see if they can also harbor the phases we discussed, in particular, investigate if the couplings in these systems are such that their ``natural" values can stabilize the phases we found.

\section{Acknowledgements}
GM, ACB, and UK are grateful to the International Centre for Theoretical Sciences (ICTS) for supporting the program - Condensed Matter meets Quantum Information (code: ICTS/COMQUI2023/9), where parts of this project were conceived. GM is also grateful to ICTS for its hospitality in summer 2024 when this work was being completed, and to the VAJRA scheme of SERB, Government of India, for its support under grant number VJR/2017/000114. ACB acknowledges the ICTS for partly supporting this research via the program - Engineered 2D Quantum Materials (code: ICTS/E2QM2024/07). U.K. is supported by the Resnick fellowship from the Bar-Ilan University, Israel, and by a fellowship from the Israel Science Foundation (ISF) grant No.~993/19. J.A. is grateful to the University of Kentucky Center for Computational Sciences and Information Technology Services Research Computing for using the Morgan Compute Cluster. Some computational portions of this work were undertaken on the Nandadevi and Kamet supercomputers, maintained and supported by the Institute of Mathematical Science's High-Performance Computing Center. Exact diagonalization calculations were performed using the DiagHam libraries~\cite{DiagHam}. ACB thanks the Science and Engineering Research Board (SERB) of the Department of Science and Technology (DST) for funding support via the Mathematical Research Impact Centric Support (MATRICS) Grant No. MTR/2023/000002.

\bibliography{hall}

\begin{appendices}

\section{Hartree-Fock form of the variational energy}
\label{app: A}
It is surprising that despite the strongly correlated nature of the states, we can write variational energy in a HF-like form. In this appendix, we will show why this is true for the states we consider. 

We reproduce the expression for the variational energy, Eq.~\eqref{energy_functional}.

\bean\label{energy_functional_app}
&&\quad\frac{1}{N_\phi}\langle\Psi|\hat H_{an}|\Psi\rangle=E_{II}+E_{IF}+
E_{FF},\nn
&&E_{II}\sim\langle I|\hat C^\dagger_I\hat C^\dagger_I\hat C_I\hat C_I|I\rangle,\ E_{IF}\sim\langle I|\hat C^\dagger_I\hat C^\dagger_F\hat C_I\hat C_F|F\rangle,\nn
&&E_{FF}\sim\langle F|\hat C^\dagger_F\hat C^\dagger_F\hat C_F\hat C_F|F\rangle,
\eean 

Let us deal with each of the contributions in turn. $E_{II}$ obviously has a HF form because it deals only with fully filled spinors, so we will not consider it further. 

Let us consider $E_{IF}$. Schematically, the nonzero contributions will only come from terms $\sim \hat C^\dagger_I\hat C^\dagger_F\hat C_I\hat C_F $.  Writing all possible terms explicitly, we find, 
\bean
E_{IF}&=&\frac{1}{2}{\mathbf U}^{(m)}_{m_1m_2m_3m_4}{\mathcal M}_{\alpha\lambda}{\mathcal M}_{\beta\eta}\times\nn
&&\Big[\langle I,F|\hat c^\dagger_{m_1,I_\alpha}\hat c^\dagger_{m_2,F_\beta}\hat c_{m_3,F_\eta}\hat c_{m_4,I_\lambda}|I,F\rangle\nn
&&+\langle I,F|\hat c^\dagger_{m_1,I_\alpha}\hat c^\dagger_{m_2,F_\beta}\hat c_{m_3,I_\eta}\hat c_{m_4,F_\lambda}|I,F\rangle\nn
&&+\langle I,F|\hat c^\dagger_{m_1,F_\alpha}\hat c^\dagger_{m_2,I_\beta}\hat c_{m_3,F_\eta}\hat c_{m_4,I_\lambda}|I,F\rangle\nn
&&+\langle I,F|\hat c^\dagger_{m_1,F_\alpha}\hat c^\dagger_{m_2,I_\beta}\hat c_{m_3,I_\eta}\hat c_{m_4,F_\lambda}|I,F\rangle\Big].\label{eq: EIF_app}
\eean

In Eq.~\eqref{eq: EIF_app} we have used ${\mathcal M}$ to represent the $4{\times}4$ anisotropic matrices $\tau_a$ appearing in the interactions. The key is to realize that in the computation of the variational energy, the states on the right and left are identical. Hence the integer-occupied particles must be created and destroyed in the same orbital, with the same being true for the fractionally occupied spinor. Hence we obtain 
\bean
&&\langle I,F|\hat c^\dagger_{m_1,I_\alpha}\hat c^\dagger_{m_2,F_\beta}\hat c_{m_3,F_\eta}\hat c_{m_4,I_\lambda}|I,F\rangle\nn
&=&\langle I|\hat c^\dagger_{m_1,I_\alpha}\hat c_{m_4,I_\lambda}|I\rangle\times\langle F|\hat c^\dagger_{m_2,F_\beta}\hat c_{m_3,F_\eta}|F\rangle\nn
&=&\delta_{m_1m_4}(P_I)_{\lambda\alpha}\times\delta_{m_2m_3}\nu_F (P_F)_{\eta\beta}.
\eean 
Using this and the fermionic commutation relation, we reach the HF-like expression for $E_{IF}$ in Eq.\eqref{eii_eif}.

Now we consider $E_{FF}$. For the convenience of the readers, we reproduce the expression for the anisotropy energy ${\cal E}^{an}$ of Eq.~\eqref{HF_form}. 
\bean\label{HF_form_app}
&&\mathcal{E}^{\rm an}(P_1,P_2,u_\perp,u_z)=\nn
&&\quad u_{\perp,H} \big[{\rm Tr}(P_1\tau_x){\rm Tr}(P_2\tau_x)+{\rm Tr}(P_1\tau_y){\rm Tr}(P_2\tau_y)\big]\nn
&&+u_{z,H}{\rm Tr}(P_1\tau_z){\rm Tr}(P_2\tau_z)\nn
&&-u_{\perp,F} \big[{\rm Tr}(P_1\tau_xP_2\tau_x)+{\rm Tr}(P_1\tau_yP_2\tau_y)\big]\nn
&&-u_{z,F}{\rm Tr}(P_1\tau_zP_2\tau_z),
\eean
For spinor-polarized FQHE states, there is only one fractionally filled spinor, and $P_1{=}P_2{=}P_F$ is a projector to a one-dimensional subspace. The orbital structure determines the averages of the Haldane pseudopotentials. This orbital structure is identical to the fully polarized FQH state in semiconductor heterostructures and is computed by exact diagonalization. Now we note the following two facts: (i) Only odd $m$ Haldane pseudopotentials have nonvanishing averages in a one-dimensional spinor subspace. For odd $m$ we also have $u_{iH}{=}{-}u_{iF}$. (ii) When projected to a one-dimensional subspace, the matrices $\tau_a$ become pure numbers. Combining these two facts we see that the expression for ${\cal E}^{an}$ is simply the strength of the odd-$m$ pseudopotential, and is the correct answer for the 2CSs.\\

Finally, we look at singlet 3CSs. Here there are two partially occupied spinors. Let us schematically denote them as $\Uparrow$ and $\Downarrow$, keeping in mind that they do not represent spin, but merely two orthogonal 4-component spinors.
The projector to the fractional subspace is $P_F{=}|\Uparrow\rangle\langle\Uparrow|{+}|\Downarrow\rangle\langle\Downarrow| $. Now we project the anisotropic matrices ${\mathcal M}$ to this subspace. 
\bean
\tilde{\mathcal M}=
\begin{pmatrix}
\langle \Uparrow |{\mathcal M}|\Uparrow\rangle & \langle \Uparrow |{\mathcal M}|\Downarrow\rangle \\
\langle \Downarrow |{\mathcal M}|\Uparrow\rangle & \langle \Downarrow |{\mathcal M}|\Downarrow\rangle 
\end{pmatrix}
=
\begin{pmatrix}
\tilde{\mathcal M}_{\Uparrow\Uparrow}  & \tilde{\mathcal M}_{\Uparrow\Downarrow}  \\
\tilde {\mathcal M}_{\Downarrow\Uparrow} & \tilde{\mathcal M}_{\Downarrow\Downarrow}
\end{pmatrix}
\eean 
Then the projected anisotropic interaction with Haldane pseudopotential $m$ will read 
\bean
\hat V_{an}^{(m)}=\tilde{\mathcal M}_{ss^\prime}\tilde{\mathcal M}_{tt^\prime}{\mathbf U}^{(m)}_{m_1m_2m_3m_4}\hat c^\dagger_{m_1s}\hat c^\dagger_{m_2t}\hat c_{m_3t^\prime}\hat c_{m_4s^\prime},\nn
\eean
where $s,t$ can take values $\Uparrow,\Downarrow$ and an implicit sum over repeated indices is assumed. 
We evaluate the expectation value  of ${\hat V}_{an}^{(m)}$ for singlet states 
\bean\label{eff}
&&\langle F|\hat V_{an}^{(m)}|F\rangle
=\tilde{\mathcal M}_{\Uparrow\Uparrow}\tilde{\mathcal M}_{\Uparrow\Uparrow} \langle V^{(m)}\rangle_{\Uparrow\Uparrow}
+\tilde{\mathcal M}_{\Downarrow\Downarrow}\tilde{\mathcal M}_{\Downarrow\Downarrow}\langle V^{(m)}\rangle_{\dwa\dwa}\nn
&&\quad\quad\quad\quad\quad\ +\biggl[\tilde{\mathcal M}_{\Uparrow\Uparrow}\tilde{\mathcal M}_{\Downarrow\Downarrow}-(-1)^m\tilde{\mathcal M}_{\Uparrow\Uparrow}\tilde{\mathcal M}_{\Downarrow\Uparrow}\biggl]\langle V^{(m)}\rangle_{\Uparrow\Downarrow},\nn
&&\langle V^{(m)}\rangle_{\Uparrow\Uparrow}=\langle F|{\mathbf U}^{(m)}_{m_1m_2m_3m_4}\hat c^\dagger_{m_1\Uparrow}\hat c^\dagger_{m_2\Uparrow}\hat c_{m_3\Uparrow}\hat c_{m_4\Uparrow}|F\rangle,\nn
&&\langle V^{(m)}\rangle_{\Downarrow\Downarrow}=
\langle F|{\mathbf U}^{(m)}_{m_1m_2m_3m_4}\hat c^\dagger_{m_1\Downarrow}\hat c^\dagger_{m_2\Downarrow}\hat c_{m_3\Downarrow}\hat c_{m_4\Downarrow}|F\rangle,\nn
&&\langle V^{(m)}\rangle_{\Uparrow\Downarrow}=
2\langle F|{\mathbf U}^{(m)}_{m_1m_2m_3m_4}\hat c^\dagger_{m_1\Uparrow}\hat c^\dagger_{m_2\Downarrow}\hat c_{m_3\Downarrow}\hat c_{m_4\Uparrow}|F\rangle.
\eean 

It is easy to see that
\bean
&&\langle V^{(m)}\rangle_{\Uparrow\Uparrow}+\langle V^{(m)}\rangle_{\Downarrow\Downarrow}+\langle V^{(m)}\rangle_{\Uparrow\Downarrow}=\langle \hat V^{(m)}\rangle\nn
&&\hat V^{(m)}=\sum_{ss^\prime}{\mathbf U}^{(m)}_{m_1m_2m_3m_4}\hat c^\dagger_{m_1s}\hat c^\dagger_{m_2s^\prime}\hat c_{m_3s^\prime}\hat c_{m_4s}.
\eean
Our exact diagonalization results show that the following relations hold for singlet states:
\begin{itemize}[left=0pt]
    \item When $m$ is odd,
    \bean
    \langle V^{(m)}\rangle_{\Uparrow\Uparrow}=\langle V^{(m)}\rangle_{\Downarrow\Downarrow}=\langle V^{(m)}\rangle_{\Uparrow\Downarrow}=\frac{1}{3}\langle  \hat V^{(m)}\rangle,
    \eean 
    which allows us to pull out an overall constant in Eq.~\eqref{eff}, leading to
     \bean
    &&\langle F|\hat V_{an}^{(m)}|F\rangle\nn
    &\propto& \biggl(\tilde{\mathcal M}_{\Uparrow\Uparrow}\tilde{\mathcal M}_{\Uparrow\Uparrow}+\tilde{\mathcal M}_{\Downarrow\Downarrow}\tilde{\mathcal M}_{\Downarrow\Downarrow}+\tilde{\mathcal M}_{\Uparrow\Uparrow}\tilde{\mathcal M}_{\Downarrow\Downarrow}+\tilde{\mathcal M}_{\Uparrow\Downarrow   }\tilde{\mathcal M}_{\Downarrow\Uparrow}\biggl)\nn
    &=&
    \frac{1}{2}tr\biggl[\tilde{\mathcal M}P_F\biggl]^2+\frac{1}{2}tr\biggl[\tilde{\mathcal M}P_F\tilde{\mathcal M}P_F\biggl].
    \eean 
    This is precisely the HF-like form [Eq.~\eqref{HF_form_app}] we desire. 
    
    \item When $m$ is even, just from fermionic antisymmetry we have (holds for even non-singlet states)
    \bean
    \langle V^{(m)}\rangle_{\Uparrow\Uparrow}=\langle V^{(m)}\rangle_{\Downarrow\Downarrow}=0,
    \quad\langle V^{(m)}\rangle_{\Uparrow\Downarrow}=\langle \hat V^{(m)}\rangle.
    \eean 
Thus, we have
    
    \bean
    &&\langle F|\hat V_{an}^{(m)}|F\rangle\nn
    &\propto& \biggl(\tilde{\mathcal M}_{\Uparrow\Uparrow}\tilde{\mathcal M}_{\Downarrow\Downarrow}-\tilde{\mathcal M}_{\Uparrow\Downarrow}\tilde{\mathcal M}_{\Downarrow\Uparrow}\biggl)\nn
    &=&\frac{1}{2}tr\biggl[\tilde{\mathcal M}P_F\biggl]^2-\frac{1}{2}tr\biggl[\tilde{\mathcal M}P_F\tilde{\mathcal M}P_F\biggl],
    \eean 
    which is also in the HF-like form.
\end{itemize}
Thus, we can express $E_{FF}$ for a singlet FQHE state as
\bean
E_{FF}=\frac{1}{2}\sum_m\frac{\langle \hat V^{(m)}\rangle}{[2-(-1)^m]N_\phi}\times\frac{1}{2}
\mathcal{E}^{\rm an}(P_F,P_Fq^{\prime},u^{(m)}_\perp,u^{(m)}_z).\nn
\eean 

Note that for the singlet states, the variational energy should be invariant under arbitrary $SU(2)$ transformations of the spinors in the fractional subspace. This places severe constraints on the form that the variational energy can take. The HF-like form, depending only on the total projector to the fractional subspace $P_F{=}P_{\Uparrow}{+}P_{\Downarrow}$, is invariant under these transformations by construction.
\section{Phase Diagrams in the ZLL}
\label{app: B}

From Eq.~\eqref{n0_haldane}, we see that when projected to the $\bn{=}0$ LL manifold, the USR interaction has only the $m{=}0$ Haldane pseudopotentials being nonvanishing, which results in 
\bean
u_{\perp,H}=u_{\perp,F}=u^{(0)}_\perp,\ u_{z,H}=u_{z,F}=u^{(0)}_z.
\eean 
The FQHE phase diagrams for different fillings with these USR couplings $u^{(0)}_\perp$ and $u^{(0)}_z$ were first studied in \cite{Sodemann_MacDonald_2014}. This makes the variational calculation simple because for nearly all FQHE states, $\langle F|\hat V^{(0)}|F\rangle{\approx} 0$. In \cite{Jincheng2024}, three of us relaxed the USR limit of the interaction by making the $m{=}1$ Haldane pseudopotentials also nonzero. Specifically, in Fourier space 
\bean
V_{ij}(q)=1+v^{(1)}_{ij}L_1(q^2\ell^2).
\eean 
Projection to ZLL  will lead to 
\bean
u^{(0)}_i=\frac{g_{i0}+g_{iz}}{4\pi\ell^2},\quad 
u^{(1)}_i=\frac{v^{(1)}_{i0}g_{i0}+v^{(1)}_{iz}g_{iz}}{4\pi\ell^2}. 
\eean 

Simplifying the notation a bit, the resulting Hartree Fock couplings will be 
\bean\label{zll_hf}
&&u_{\perp,H}=u^{(0)}_{\perp}+u^{(1)}_{\perp}=(1+v_{1\perp})\frac{g_{\perp 0}+g_{\perp z}}{4\pi\ell^2},\nn
&&u_{\perp,F}=u^{(0)}_{\perp}-u^{(1)}_{\perp}=(1-v_{1\perp})\frac{g_{\perp 0}+g_{\perp z}}{4\pi\ell^2},\nn
&&u_{z,H}=u^{(0)}_{z}+u^{(1)}_{z}=(1+v_{1z})\frac{g_{z0}+g_{zz}}{4\pi\ell^2},\nn 
&&u_{z,F}=u^{(0)}_{z}-u^{(1)}_{z}=(1-v_{1z})\frac{g_{z0}+g_{zz}}{4\pi\ell^2}.
\eean
As can be seen, $g_{i0}$ and $g_{iz}$ always appear in the combination $(g_{i0}{+}g_{iz})$. Therefore, without losing generality, we can $g_{i0}{=}0$ (which is their nominal value in the microscopic estimate~\cite{Wei_Xu_Sodemann_Huang_LLM_SU4_breaking_MLG_2024}, and build a unique map between $(g_{\perp z},v_{1\perp},g_{z z},v_{1z})$ and $(u_{\perp, H},\ u_{\perp, F},\ u_{z, H},\ u_{z, F} )$.\\
We will fix $v_{1\perp}$ and $v_{1z}$ and vary $(g_{\perp z},g_{z z})$, so that the ratio between Hartree and Fock couplings is fixed, i.e.
\bean
u_{\perp,H}/u_{\perp,F}=\frac{1+v_{1\perp}}{1-v_{1\perp}},\ 
u_{z,H}/u_{z,F}=\frac{1+v_{1z}}{1-v_{1z}}.
\eean 

Note that this corresponds to the type II phase diagrams. 
Note also that we used a different parameterization in our previous work~\cite{Jincheng2024} to explore the coupling space $(u_{\perp, H},\ u_{\perp, F},\ u_{z, H},\ u_{z, F} )$, consistent with work by Stefanidis and Sodemann on the USR case~\cite{Stefanidis_Sodemann2023} where the differences between Hartree and Fock couplings $\Delta_\perp{=}u_{\perp, H}{-}u_{\perp, F}$, $\Delta_z{=}u_{z, H}{-}u_{z, F}$ are held fixed, while $(u_{\perp, H},\ u_{z, H})$ are varied. 

Our objectives in this appendix are twofold: (i) We present phase diagrams for $\nu=-\frac{3}{5}$ for both 2CS and 3CS for natural couplings in the ZLLs. (ii) We also show that the phase diagrams in the $\bn{=}1$ LL manifold presented in the main text can be obtained in the ZLLs provided one makes the Hartree parts of the couplings very small ($u_{z, H},~u_{\perp, H}\sim 10^{-2}E_z$). This mimics the natural couplings in the $\bn{=}1$ LL manifold. As a side-benefit, we see the ZLL and $\bn=1$ phase diagrams side by side, which provide a useful contrast. Furthermore, we find that even though our ZLL Hamiltonian has only the $m{=}0,1$ Haldane pseudopotentials nonvanishing, all the phases found in the $\bn{=}1$ LL manifold, with $V_m{\neq}0,\ \ m{=}0,1,2$ are also found in the ZLL phase diagrams. This shows that the $m=2$ pseudopotential is not making a qualitative difference.

We start with the phase diagrams for the 2CS $(1,\frac{2}{3},0,0)$ in Fig.~\ref{Fig: 5} for natural couplings in the ZLLs. The corresponding phase diagrams for  $(1,\frac{2}{5},0,0)$  in the ZLLs are shown in Fig.~\ref{Fig: 11}. Recall that natural couplings in the ZLLs have the Hartree couplings large.  The ratio of the $m{=}1$ to $m{=}0$ couplings is kept fixed while the $m{=}0$ coupling is varied. Note that we see the BOFM phase in a large region of the coupling constants. In our earlier work~\cite{Jincheng2024} on phase diagrams for $(1,\frac{2}{3},0,0)$ in the ZLLs, we never saw the BOFM phase. The reason is that for the fraction $(1,\frac{2}{3},0,0)$, the numerical coefficients are such that the VPFM and BOFM phases are exactly degenerate for $E_v=0$. Since in real samples in the ZLLs, there is always some nonzero, albeit tiny, $E_v$, the VPFM phase always wins for $(1,\frac{2}{3},0,0)$. This is not true in the case of $(1,\frac{2}{5},0,0)$, for which the numerical coefficients in front of the various pseudopotentials do not result in an exact degeneracy between the VPFM and BOFM phases. Now there is a true competition between the phases, resulting in a large region of BOFM in the phase diagram. Additionally, we did not see the BOCAFM or BOAFM phases in our previous work~\cite{Jincheng2024}, which appear in small regions of Fig.~\ref{Fig: 11}. We attribute this to the different parameterization used in that work, which makes the regions over which these two phases are stable very tiny, and close to the origin. \\ 
In Fig.~\ref{Fig: 12} we show the phase diagrams for the same fraction, but now with natural couplings adjusted to be in the $\bn\neq0$ LL manifold. Recall that this implies that the Hartree couplings are fixed, and much smaller than the Fock couplings. One can easily see the huge difference in the topology of the phase diagram. The same phases now appear in completely different quadrants. Furthermore, the phase diagrams are very similar to those of type II phase diagrams in Fig.~\ref{Fig: 2}, though with some differences. We attribute these differences to the fact that here we have set the $m=2$ pseudopotentials to zero, whereas in Fig.~\ref{Fig: 2} they were nonzero, and prescribed by the USR couplings $g_{ij}$. 

\begin{figure*}[h]
    \centering
  \includegraphics[width=1.05\textwidth,height=0.25\textwidth]{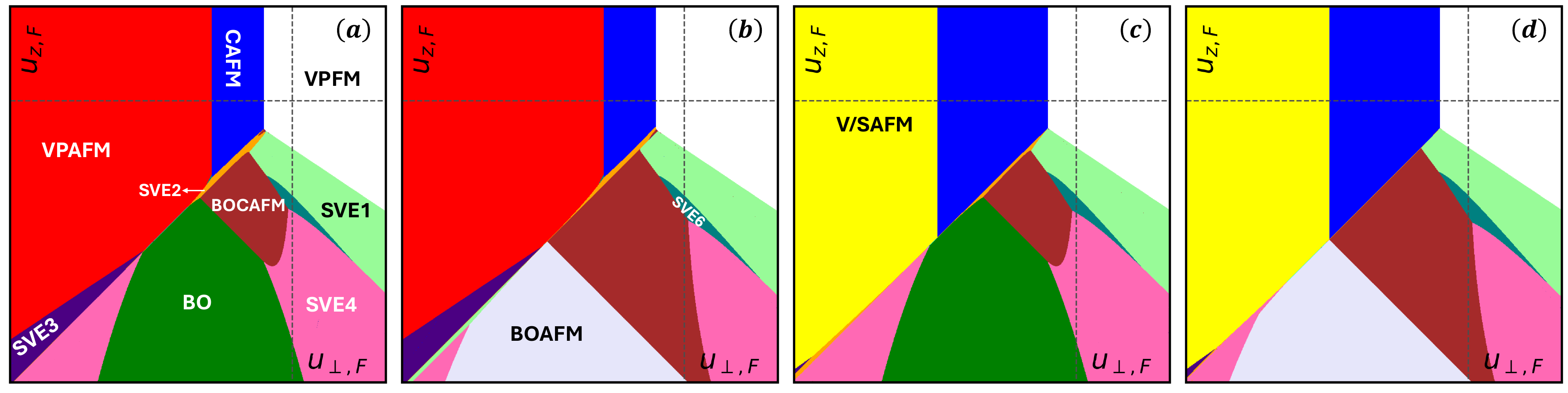}
    \caption{ $(1,\frac{2}{3},0,0)$  phase diagrams with $u_{\perp,F}\in[-3,1]$ and $u_{z,F}\in[-3,1]$ in unit of $E_z$. The Hartree couplings are very small ($\sim 10^{-2}E_z$) and fixed, to mimic the natural couplings in the $\bn\neq0$ LL manifolds. (a) $u_{\perp,H}<0,\ u_{z,H}<0$, (b) $u_{\perp,H}<0,\ u_{z,H}>0$, (c) $u_{\perp,H}>0,\ u_{z,H}<0$, (d) $u_{\perp,H}>0,\ u_{z,H}>0$.}
    \label{Fig: 13}
\end{figure*}

\begin{widetext}

\begin{figure}[H]
    \centering
  \includegraphics[width=1.05\textwidth,height=0.25\textwidth]{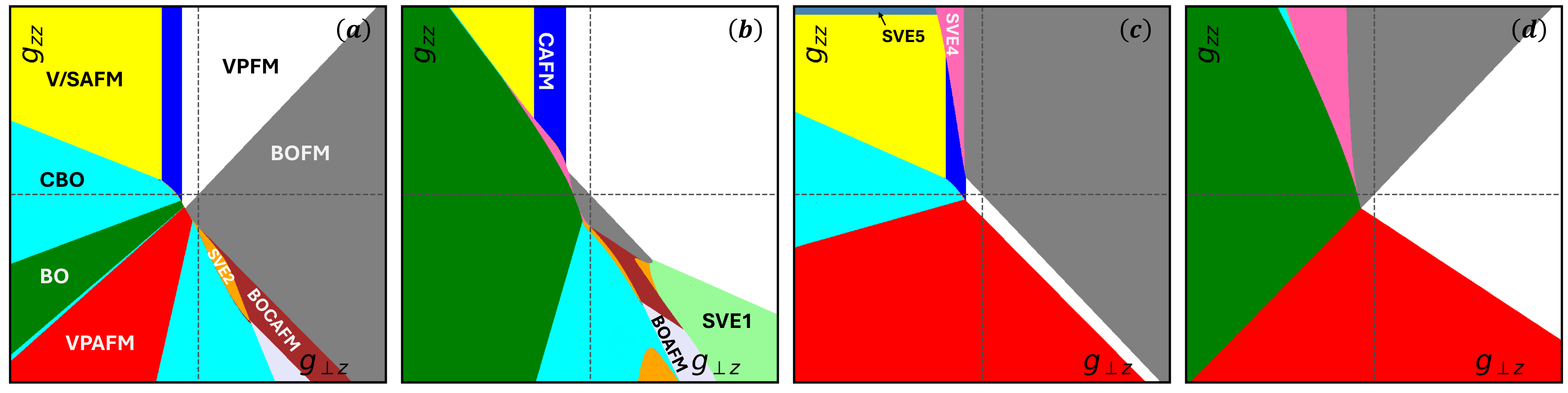}
    \caption{Phase diagrams  for the filling $(1,\frac{2}{5},0,0)$ in the ZLLs with $ g_{\perp z}\in [-1600,1600]\text{meV}\cdot \text{nm}^{2}$ and $ g_{zz}\in [-1600,1600]\text{meV}\cdot \text{nm}^{2}$. The $m=1$ Haldane pseudopotentials of each $g_{ij}$ are $0.2$ times the corresponding $m=0$ pseudopotentials in magnitude. (a) $u^{(1)}_\perp=-0.2u^{(0)}_\perp,~u^{(1)}_z=-0.2u^{(0)}_z$. (b) $u^{(1)}_\perp=0.2u^{(0)}_\perp,~u^{(1)}_z=-0.2u^{(0)}_z$. (c) $u^{(1)}_\perp=-0.2u^{(0)}_\perp,~u^{(1)}_z=0.2u^{(0)}_z$. (d) $u^{(1)}_\perp=0.2u^{(0)}_\perp,~u^{(1)}_z=0.2u^{(0)}_z$. 
}
    \label{Fig: 11}
\end{figure}

\begin{figure}[H]
    \centering
  \includegraphics[width=1.05\textwidth,height=0.25\textwidth]{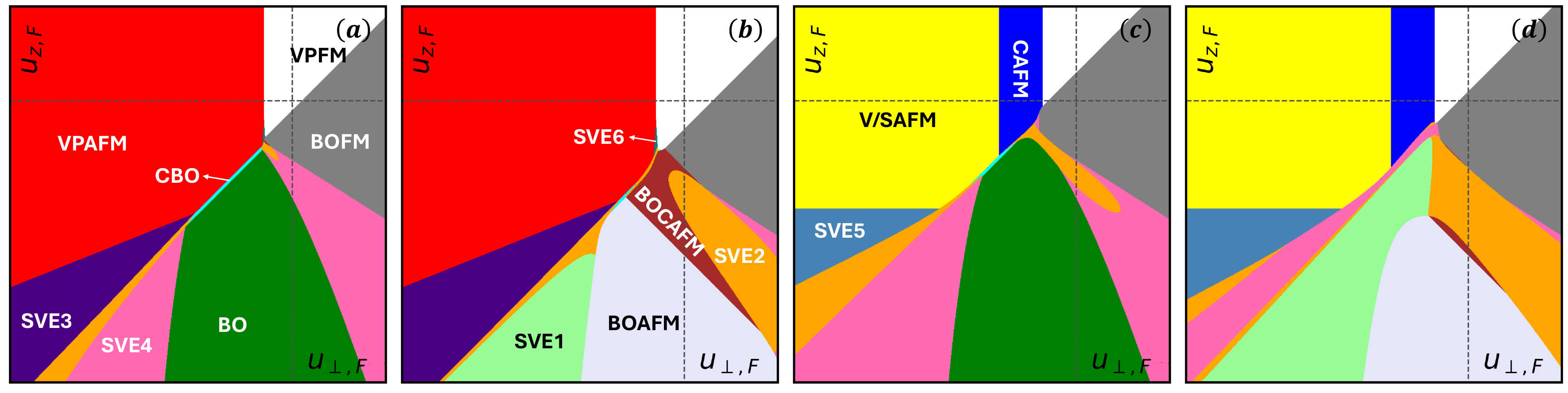}
    \caption{ $(1,\frac{2}{5},0,0)$  phase diagrams with $u_{\perp,F}\in[-3,1]$ and $u_{z,F}\in[-3,1]$ in unit of $E_z$. The Hartree couplings are very small ($\sim 10^{-2}E_z$) and fixed, to mimic the natural couplings in the $\bn\neq0$ LL manifolds. (a) $u_{\perp,H}<0,\ u_{z,H}<0$, (b) $u_{\perp,H}<0,\ u_{z,H}>0$, (c) $u_{\perp,H}>0,\ u_{z,H}<0$, (d) $u_{\perp,H}>0,\ u_{z,H}>0$.}
    \label{Fig: 12}
\end{figure}
\end{widetext}
In Fig.~\ref{Fig: 13}, we present the phase diagrams for $(1,\frac{2}{3},0,0)$ with very small Hartree couplings, mimicking the natural couplings in the $\bn{\neq}0$ LL manifolds. As can be seen, the phase diagrams are very similar to the type II phase diagrams presented in Fig.~\ref{Fig: 4}, once again with some quantitative differences because the $m{=}2$ pseudopotentials are set to zero here.  

\begin{figure*}[h]
    \centering
  \includegraphics[width=1.05\textwidth,height=0.25\textwidth]{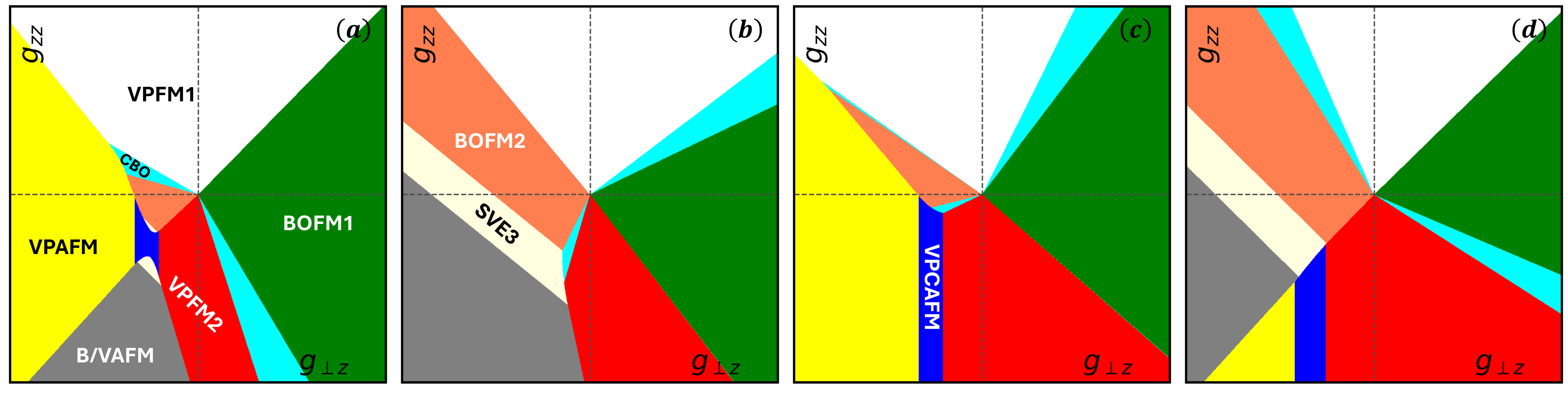}
    \caption{Type II section phase diagrams for $(1,[\frac{1}{5},\frac{1}{5}],0)$  in the ZLLs,  with $ g_{\perp z}\in [-800,800]\text{meV}\cdot \text{nm}^{2}$ and $ g_{zz}\in [-800,800]\text{meV}\cdot \text{nm}^{2}$. The $m=1$ Haldane pseudopotentials of each $g_{ij}$ are $0.1$ times the corresponding $m=0$ pseudopotentials in magnitude. (a) $u^{(1)}_\perp=-0.1u^{(0)}_\perp,~u^{(1)}_z=-0.1u^{(0)}_z$. (b) $u^{(1)}_\perp=0.1u^{(0)}_\perp,~u^{(1)}_z=-0.1u^{(0)}_z$. (c) $u^{(1)}_\perp=-0.1u^{(0)}_\perp,~u^{(1)}_z=0.1u^{(0)}_z$. (d) $u^{(1)}to^{(0)}_\perp,~u^{(1)}_z=0.1u^{(0)}_z$. }
    \label{Fig: 14}
\end{figure*}

Now we turn to the phase diagrams for the singlet 3CS. In Fig.~\ref{Fig: 14} we show the phase diagrams for the filling $(1,[\frac{1}{5},\frac{1}{5}],0)$ for natural couplings in the ZLLs (large Hartree parts of both couplings). The phases found are the same as those found in our earlier work~\cite{Jincheng2024} for $(1,[\frac{1}{3},\frac{1}{3}],0)$ with a few quantitative differences. Note that our nomenclature for the states was different in that work. The FM in the earlier work corresponds to VPFM1 here, AFMCDW1 corresponds to VPFM2 here, BO1 corresponds to BOFM1 here, AFMCDW2 corresponds to VPAFM here, and VFM corresponds to B/VAFM here. 

\begin{figure*}[h]
    \centering
  \includegraphics[width=1.05\textwidth,height=0.25\textwidth]{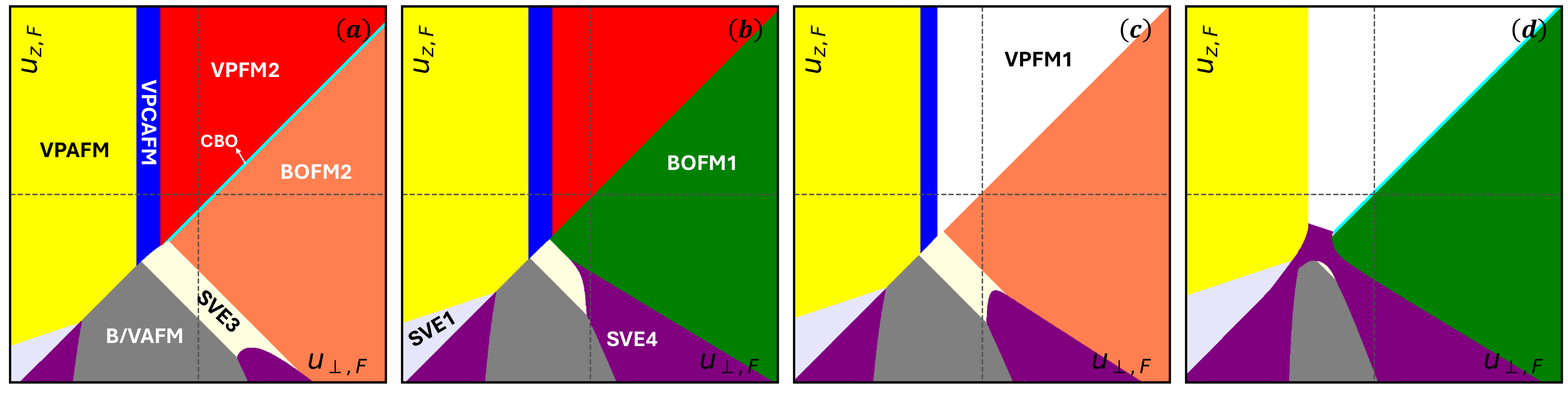}
    \caption{ $(1,[\frac{1}{5},\frac{1}{5}],0)$  with $u_{\perp,F}\in[-2,2]$ and $u_{z,F}\in[-2,2]$ in unit of $E_z$. The Hartree couplings are very small ($\sim 10^{-2}E_z$) and fixed, in order to mimic the natural couplings in the $\bn\neq0$ LL manifolds. (a) $u_{\perp,H}<0,\ u_{z,H}<0$, (b) $u_{\perp,H}<0,\ u_{z,H}>0$, (c) $u_{\perp,H}>0,\ u_{z,H}<0$, (d) $u_{\perp,H}>0,\ u_{z,H}>0$.}
    \label{Fig: 15}
\end{figure*}

\begin{figure*}[h]
    \centering
  \includegraphics[width=1.05\textwidth,height=0.25\textwidth]{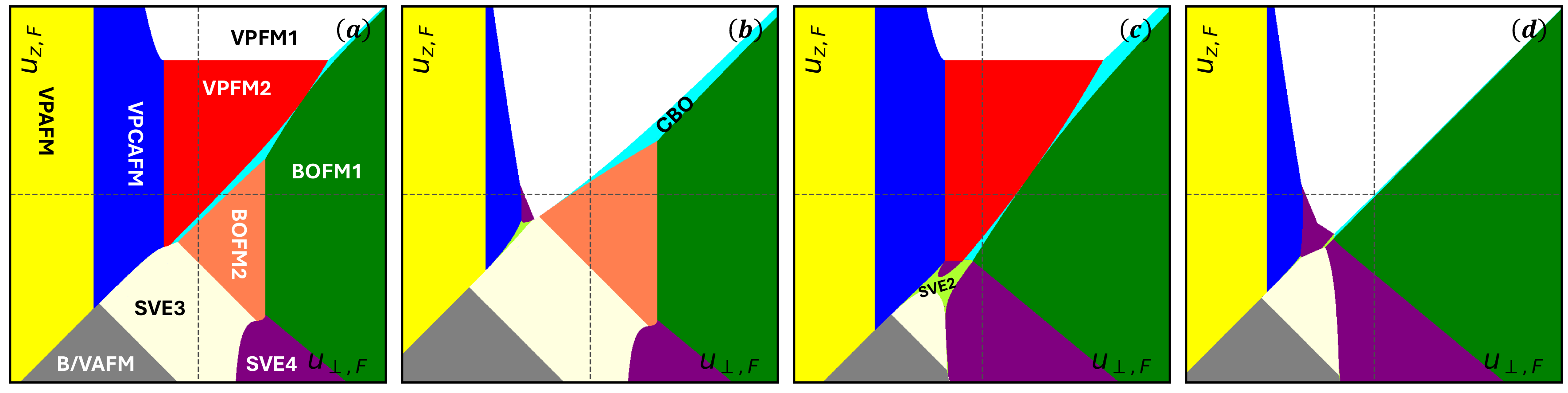}
    \caption{ $(1,[\frac{1}{3},\frac{1}{3}],0)$  phase diagrams with $u_{\perp,F}\in[-2,2]$ and $u_{z,F}\in[-2,2]$ in unit of $E_z$. The Hartree couplings are very small ($\sim 10^{-2}E_z$) and fixed, to mimic the higher LL manifolds. (a) $u_{\perp,H}<0,\ u_{z,H}<0$, (b) $u_{\perp,H}<0,\ u_{z,H}>0$, (c) $u_{\perp,H}>0,\ u_{z,H}<0$, (d) $u_{\perp,H}>0,\ u_{z,H}>0$.}
    \label{Fig: 16}
\end{figure*}

In Fig.~\ref{Fig: 15} we mimic the natural couplings in the $\bn\neq0$ LL manifolds by fixing the Hartree parts of the couplings to be very small, and varying the Fock parts. As can be seen, the phase diagrams are quite different from those for natural couplings in the ZLLs, and very similar to those found in the main text (Fig.~\ref{Fig: 7}) for USR $g_{ij}$ in the $\bn=1$ LL manifold. Similarly, in Fig.~\ref{Fig: 16} we mimic the natural couplings in the $\bn\neq0$ LL manifolds for the filling $(1,[\frac{1}{3},\frac{1}{3}],0)$. The results are very similar to Fig.~\ref{Fig: 10} in the main text, despite the absence of the $m=2$ pseudopotentials here. 
 
\begin{figure*}[h]
    \centering
  \includegraphics[width=1.05\textwidth,height=0.25\textwidth]{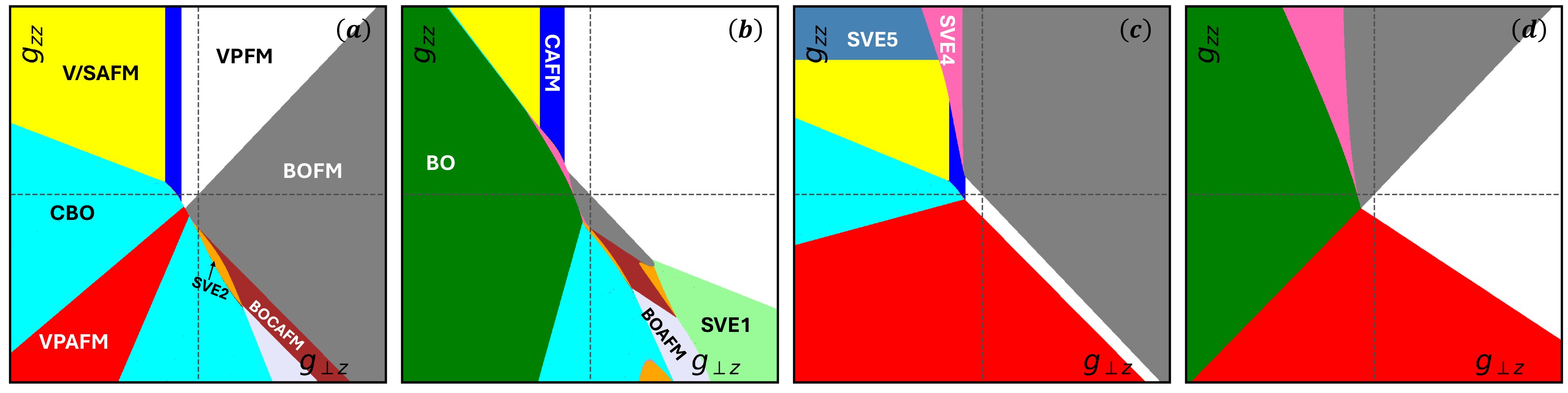}
    \caption{Phase diagrams  for the filling $(1,\frac{1}{3},0,0)$ in the ZLLs with $ g_{\perp z}\in [-1600,1600]\text{meV}\cdot \text{nm}^{2}$ and $ g_{zz}\in [-1600,1600]\text{meV}\cdot \text{nm}^{2}$. The $m=1$ Haldane pseudopotentials of each $g_{ij}$ are $0.2$ times the corresponding $m=0$ pseudopotentials in magnitude. (a) $u^{(1)}_\perp=-0.2u^{(0)}_\perp,~u^{(1)}_z=-0.2u^{(0)}_z$. (b) $u^{(1)}_\perp=0.2u^{(0)}_\perp,~u^{(1)}_z=-0.2u^{(0)}_z$. (c) $u^{(1)}_\perp=-0.2u^{(0)}_\perp,~u^{(1)}_z=0.2u^{(0)}_z$. (d) $u^{(1)}_\perp=0.2u^{(0)}_\perp,~u^{(1)}_z=0.2u^{(0)}_z$. 
}
    \label{Fig: 17}
\end{figure*}

Now we turn to the filling $(1,\frac{1}{3},0,0)$. In Fig.~\ref{Fig: 17} we show the phase diagrams for this filling for natural couplings in the ZLLs. The phase diagrams are quite similar to the $(1,\frac{2}{5},0,0)$ phase diagrams (Fig.~\ref{Fig: 11}).

\begin{figure*}[h]
    \centering
  \includegraphics[width=1.05\textwidth,height=0.25\textwidth]{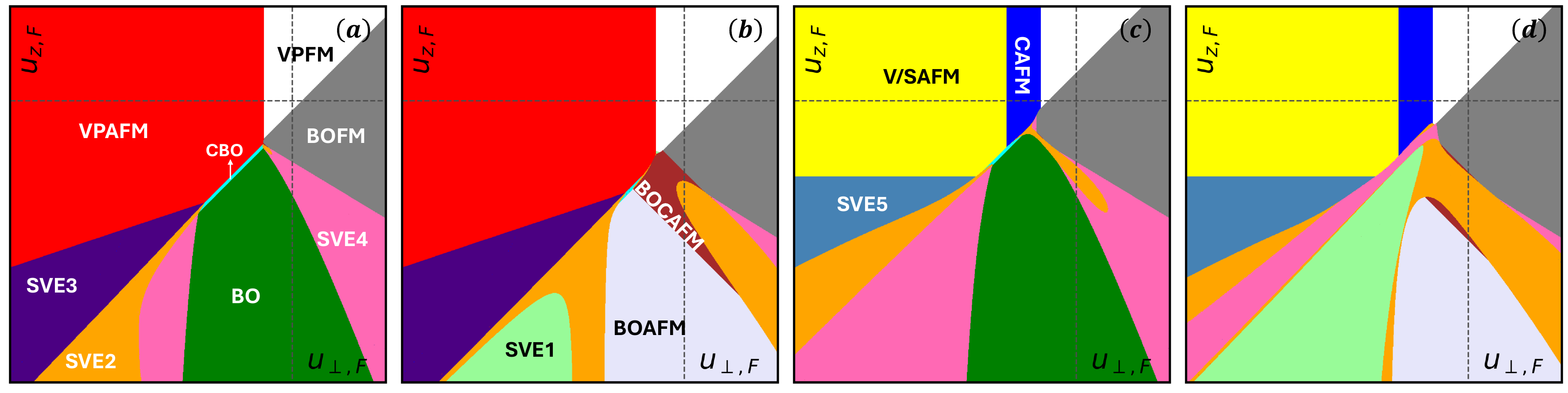}
    \caption{ $(1,\frac{1}{3},0,0)$  phase diagrams with $u_{\perp,F}\in[-3,1]$ and $u_{z,F}\in[-3,1]$ in unit of $E_z$. The Hartree couplings are very small ($\sim 10^{-2}E_z$) and fixed, to mimic the higher LL manifolds. (a) $u_{\perp,H}<0,\ u_{z,H}<0$, (b) $u_{\perp,H}<0,\ u_{z,H}>0$, (c) $u_{\perp,H}>0,\ u_{z,H}<0$, (d) $u_{\perp,H}>0,\ u_{z,H}>0$.}
    \label{Fig: 18}
\end{figure*}

To finish this appendix, we present the phase diagrams for the type II section for $(1,\frac{1}{3},0,0)$ with couplings appropriate to $\bn\neq0$ LL manifolds in Fig.~\ref{Fig: 18}. As one can see, the topology of the phase diagrams is considerably different from those in the ZLLs.

\end{appendices}

\end{document}